%% file: TWC_GT_arXiv_v3.tex
\documentclass[journal,12pt,onecolumn,draftclsnofoot]{IEEEtran}
\usepackage{cite}
\usepackage{graphicx,color,epsfig,rotating}
\usepackage{amsfonts,amsmath,amssymb,bbm}
\usepackage{algorithm} 
\usepackage{algpseudocode} 
\usepackage{subfigure}
\usepackage{amsmath}
\usepackage{cite}
\usepackage{placeins}
\usepackage{graphicx}
\usepackage[latin1]{inputenc}
\usepackage{amssymb}
\usepackage{multirow}
\usepackage{stfloats}
\usepackage{tabularx} 
\usepackage{booktabs} 
\usepackage{url}
\usepackage{bm}
\usepackage{soul}
\usepackage{pstricks}
\usepackage{algorithm, setspace}
\usepackage{caption}
\usepackage{lipsum}

\input{macros}

\newtheorem{defn}{Definition}

\newtheorem{theorem}{Theorem}
\newtheorem{lemma}{Lemma}

\newtheorem{proposition}{Proposition}

\newtheorem{remark}{Remark}

\makeatletter
\newcommand{\subalign}[1]{
  \vcenter{%
    \Let@ \restore@math@cr \default@tag
    \baselineskip\fontdimen10 \scriptfont\tw@
    \advance\baselineskip\fontdimen12 \scriptfont\tw@
    \lineskip\thr@@\fontdimen8 \scriptfont\thr@@
    \lineskiplimit\lineskip
    \ialign{\hfil$\m@th\scriptstyle##$&$\m@th\scriptstyle{}##$\crcr
      #1\crcr
    }%
  }
}
\makeatother

 \ifodd 1

 \else

 \fi

\ifodd 1

\else

\fi

\begin{document}
\title{A Non-cooperative Game-based 
Distributed Beam Scheduling Framework for 5G Millimeter-Wave Cellular Networks} 
\author{Xiang~Zhang,~\IEEEmembership{Student Member,~IEEE},
        Shamik Sarkar,~\IEEEmembership{Student Member,~IEEE},
         Arupjyoti Bhuyan,~\IEEEmembership{Senior Member,~IEEE},
        Sneha Kumar Kasera,~\IEEEmembership{Senior Member,~IEEE},
        and~Mingyue Ji,~\IEEEmembership{Member,~IEEE}
\thanks{Part of this work \cite{zhang2020stochastic} was presented in the IEEE Asilomar conference 2020, Pacific Grove, CA, USA.}
\thanks{X. Zhang and M. Ji are with the Department of Electrical and Computer Engineering, University of Utah, Salt Lake City, UT 84112, USA (E-mail:~\{xiang.zhang, mingyue.ji\}@utah.edu).} 
\thanks{S. Sarkar and S. K. Kasera are with the School of Computing, University of Utah, Salt Lake City, UT 84112, USA (E-mail: shamik.sarkar@utah.edu, kasera@cs.utah.edu).}
\thanks{A. Bhuyan is with the INL Wireless Security Institute, Idaho National Laboratory, Idaho Falls, ID 83415 (E-mail:
arupjyoti.bhuyan@inl.gov).}
}

\maketitle
\begin{abstract}
This paper studies the problem of distributed beam scheduling for 5G millimeter-Wave (mm-Wave) cellular networks where base stations (BSs) belonging to different operators share the same spectrum without centralized coordination among them. Our goal is to design efficient distributed scheduling algorithms to maximize the network utility, which is a function of the achieved throughput by the user equipment (UEs), subject to the average and instantaneous power consumption constraints of the BSs. We propose a Media Access Control (MAC) and a power allocation/adaptation mechanism utilizing the Lyapunov stochastic optimization framework and non-cooperative games. In particular, we first decompose the original utility maximization problem into two sub-optimization problems for each time frame, which are a convex optimization problem and a non-convex optimization problem, respectively. By formulating the distributed scheduling problem as a non-cooperative game where each BS is a player attempting to optimize its own utility, we provide a distributed solution to the non-convex sub-optimization problem via finding the Nash Equilibrium (NE) of the game whose weights are determined optimally by the Lyapunov optimization framework. Finally, we conduct simulation under various network settings to show the effectiveness of the proposed game-based beam scheduling algorithm in comparison to that of several reference schemes. 
\end{abstract}

\section{Introduction} 
The use of millimeter-Wave (mm-Wave) frequencies in 5G cellular networks makes additional spectrum available and contributes directly to orders of magnitude increase in 
throughput. However, the 
available licensed spectrum for commercial mobile services is still limited even for mm-Wave frequencies. One way to address this limitation is via spectrum sharing or spectrum pooling \cite{boccardi2016spectrum,gupta2016feasibility},
by enabling the secondary utilization of additional unlicensed or shared spectrum available for use in 5G~\cite{giordani2016multi}, allowing multiple service providers to use the same frequency band.
The characteristics of the mm-Wave frequency channel and directional beamforming have a major impact on the use of spectrum pooling. It has been shown that both orthogonal pooling, where frequency bands are allocated dynamically but exclusively to one operator at a time, and non-orthogonal pooling, where the same frequency bands can be allocated to multiple operators, have significant capacity gains \cite{jorswieck2014spectrum}. Due to the high directional gain of the mm-Wave beams, concurrent transmissions are possible using the same frequency band. However, concurrent transmissions over the shared frequency bands potentially causes severe interference among operators if there is no proper
coordination.

To handle interference and improve system throughput, two different approaches have been commonly considered. In one line of research \cite{7248501,7536929,9120704,8017481,8978709,9013827,8600015}, centralized beam and transmission scheduling was studied. In \cite{7248501}, the beam alignment versus throughput trade-off was considered and a joint optimization for beam width selection and scheduling was proposed to maximize the effective system throughput. To perform the proposed scheduling algorithm, the  knowledge of the system topology  and interference are required. The authors in \cite{7536929} considered an urban 
Non-Line-of-Sight (NLOS) mm-Wave cellular system where cooperation among subsets of base stations are allowed and proposed a scheduling mechanism to handle inter-cell interference and maximize system throughput with the consideration of fairness. In addition, the authors in \cite{8017481} focused on the joint optimization of analog beam selection and user scheduling based on limited CSI and developed two codebook-based methods. 

Another line of research \cite{7959165,7794615,8815404,8107556,ning2019interference,candogan2013near,candogan2010near,sarkar2020enabling,sarkar2021uncoordinated} considered distributed approaches where there is no coordination among multiple network entities. Though distributed scheduling suffers from performance degradation compared to centralized approaches, it usually requires lower implementation complexity. In addition, distributed approaches have the advantage of enhancing system security since it is more robust to network attacks than the centralized case where 
the failure of the central controller will lead to the shutdown of the entire system. In \cite{7794615}, a multi-RAT system was considered where 5G cellular BSs co-exist with existing networks (e.g., WiGig) and share frequency bands. A co-existence mechanism was proposed in which 5G and WiGig BSs schedule their own UEs in a distributed manner to optimize their own utilities. In \cite{8815404}, a two-stage schedule-and-align scheme was proposed to facilitate efficient communication in a scenario where a BS communicates with multiple UEs through a number of distributed remote radio units with mm-Wave antennas. Moreover, non-cooperative game-based approaches \cite{980097,pang2010design,scutari2008asynchronous,5447064,ning2019interference,candogan2013near,candogan2010near} were introduced for distributed scheduling in general cellular networks, which are also suitable for mm-Wave networks. The authors in \cite{980097} first formulated the CDMA uplink power control problem as a non-cooperative game where the UEs are the  'selfish' players trying to maximize their own individual payoff via transmit power control without collaborating with others. The Nash Equilibrium of the game is proved to exist and a corresponding parallel updating algorithm was proposed to solve the equilibrium based on the measurements of the interference and noise at each UE. In \cite{pang2010design}, a cognitive radio system, consisting of a set of primary and secondary users sharing the same frequency band, was considered. Each secondary user aims to maximize its own throughput while the aggregate interference caused by these users to the primary users  should not exceed some pre-defined thresholds. The authors formulated this scheduling problem as a non-cooperative game and proved the existence and uniqueness of the equilibrium. Also, several updating algorithms were developed and shown to converge under certain conditions. \cite{ning2019interference} studied the interference management problem in mm-Wave cellular networks in the presence of Device-to-Device (D2D) transmissions. A Stackelberg game-based interference control mechanism was proposed to optimize the D2D throughout while suppressing its interference to the mm-Wave network. In \cite{candogan2013near}, the distributed power allocation problem in  a multi-cell CDMA network was considered and a {\it potential game}-based approach was proposed to provide an approximate solution to the original non-cooperative power allocation game. It was shown that by properly selecting the pricing factors of the potential game, it can converge to the unique equilibrium which is a globally optimal power allocation. This provides a good solution to the original game in the high SINR regime. Moreover, a systematic way to find the proper potential game for any underlying power allocation game was developed in \cite{candogan2010near}.

In this paper, we consider the downlink beam scheduling problem for mm-Wave cellular networks in a realistic scenario where the base stations (BSs) may belong to different operators, both private (e.g., Nokia private LTE network) and commercial (e.g., AT\&T, Verizon), and these operators share spectrum but do not cooperate with each other. In this case, distributed beam scheduling must be performed for the downlink data transmission from the BSs of different operators to the UEs. 
One advantage of the considered non-cooperative network setting lies in its security and robustness aspects because a central controller is usually vulnerable to malicious attacks. Our goal is to design efficient distributed MAC strategies together with adaptive power control to handle inter-cell interference due to spectrum sharing and to maximize the network utility as a function of the time averaged throughput of the UEs. The major novelties of this paper are as follows. First, designing adaptive distributed beam scheduling algorithms for non-cooperative operators in mm-Wave networks has not been considered by any existing work
according to our knowledge. Second, we propose a concrete approach to solve the distributed beam scheduling problem with theoretical optimality guarantee compared to heuristic solutions in the literature. 
The main contributions of this paper are summarized as follows.

\begin{itemize}
\item We first propose a novel problem formulation based on the 
Lyapunov stochastic optimization framework 
given the underlying MAC protocols (e.g., $p$-persistent, CSMA/CA) 
but with optimizable parameters (e.g., BS transmit powers). Given the average and peak power constraints of the BSs, the proposed network utility optimization problem can be decomposed into two sub-optimization problems. 
It can be shown that solving the two sub-problems in  each time frame will yield a network utility within an 
additive gap to that obtained by solving the original optimization problem. The first sub-problem is convex and involves a set of auxiliary variables which can be 
solved distributedly. The second sub-problem involves the power allocation for the UEs associated with each BS, and is stochastic and non-convex.
\item In order to solve the second sub-problem in a distributed manner, we 
formulate the scheduling problem as
a non-cooperative game in which the BSs are the players which do not cooperate with each other. Each BS has its own payoff function which is defined as a weighted sum of the total throughout achieved by the UEs associated with that BS, plus a power consumption penalization term. Surprisingly, the weights in the payoff function are optimally determined by the decomposition of the Lyapunov optimization, i.e, the parameters in the two sub-problems.
Under this game theoretic formulation, the above sub-problems can be (approximately) solved in a distributed manner by solving the Nash Equilibrium (NE) of the corresponding non-cooperative game.
\item We identify several key properties of the formulated game and propose an iterative update algorithm to compute the equilibrium. In particular, we show that the power allocation game always admits at least one pure-strategy equilibrium and provide sufficient conditions guaranteeing the uniqueness of the equilibrium. To solve the NE, we propose a parallel updating algorithm which is proved to globally converge. This parallel updating algorithm is performed periodically to provide approximate solutions to the sub-problems at each epoch. Numerical evaluation is also conducted to verify the effectiveness of the proposed game-based scheduling compared to other 
MAC protocols with optimized transmit powers.
\end{itemize}

The rest of this paper is organized as follows. Section \ref{Section: Problem Formulation} introduces the system model and formulates the network utility maximization problem. The main results are presented in Section \ref{Section: Proposed Approach}.
In Section \ref{Section: Numerical Evaluation}, a thorough numerical evaluation is provided to justify the performance of the proposed scheduling algorithm. We conclude the paper in Section~\ref{sec: Conclusion}.

\paragraph*{Notation}
Let $\mathbb{Z}^+$ denote the set of positive integers. 
Let $[n]\eqdef\{1,\cdots,n\}$ for some $n\in \mathbb{Z}^+$. For a set of real numbers $a_i,i\in[n]$, we let $(a_i)_{i=1}^{n}\eqdef [a_1,\cdots ,a_n]^{\sf T}$. $\mathbf{0}^{n}\eqdef [0,0,\cdots,0]$ denotes the all-zero row vector with $n$ entries.  Calligraphic letters $\mathcal{A},\mathcal{B},\cdots$ represent sets, bold capital letters $\mathbf{A},\mathbf{B},\cdots$ represent matrices. For a matrix $\mathbf{A}\eqdef[a_{i,j}]\in\mathbb{R}^{m\times n}$, the Frobenius norm is defined as $\|\mathbf{A}\|_2\eqdef \sqrt{\sum_{i=1}^m\sum_{j=1}^n|a_{i,j}|^2}$. For two sets $\mathcal{A}$ and $\mathcal{B}$, the \emph{difference set} is defined as $\mathcal{A}\backslash \mathcal{B}\eqdef\{x\in\mathcal{A}:x\notin \mathcal{B}\}$. We denote the Euclidean projection of $x$ onto the interval $[a,b]$ as $[x]_a^b$, i.e., $[x]_a^b=x$ if $a\le x\le b$, $[x]_a^b=a$ if $x<a$ and $[x]_a^b=b$ if $x> b$. 
All logarithms used in this paper are natural logarithm.

\section{Problem Formulation}
\label{Section: Problem Formulation}
\subsection{Network Model}
\label{Subsection: Network Model}
We consider a cellular network with $M$ base stations (BSs) and $K$ user equipments (UEs).
Each BS $i\in[M]$ belonging to an operator is responsible for serving a set of $K_i$ UEs denoted by $\mathcal{K}_i\subseteq [K]$, i.e., the UEs in $\mathcal{K}_i$ are associated with BS $i$. The total number of UEs is equal to $K=\sum_{i=1}^MK_i$. BSs from multiple operators are allowed to be co-located at the same sites. The system operates on a shared frequency band with bandwidth $W$ Hz with a center frequency at $W_{\rm c}$ Hz.  We are interested in the  downlink data transmission and scheduling for this network. Due to the proximity of locations, UEs may suffer from the interference caused by neighboring BSs of different operators. The received Signal-to-Interference-plus-Noise Ratio ({SINR}) at UE $j\in [K]$ is given by 
\be 
\textrm{SINR}_{j,i(j)}=\frac{p_{j,i(j)}G_{j,i(j)}^{\rm UE}G_{j,i(j)}^{\rm BS}|h_{j,i(j)}|^2d_{j,i(j)}^{-\eta}   }{\sum_{\ell \in\mathcal{B}(j)\backslash\{i(j)\}}p_{j(\ell),\ell}G_{j,\ell}^{\rm UE}G_{j,\ell}^{\rm BS}|h_{j,\ell}|^2d_{j,\ell}^{-\eta}    +\sigma^2}\;,
\ee 
where $i(j)\in [M]$ denotes the BS index which is transmitting to UE $j$\footnote{{For any UE $j$, we let $i(j)$ denote the BS that this UE is associated with, i.e., $j\in\mathcal{K}_{i(j)}$. Similarly, we let $j(i)\in\mathcal{K}_i$ denote the UE that is selected by BS $i$ to transmit to  (if one UE is scheduled at a time for each BS).}};
$p_{j,i(j)}, h_{j,i(j)}$ and $d_{j,i(j)}$ denote the transmit power, 
channel gain and distance from BS $i(j)$ to UE $j$, respectively. $\mathcal{B}(j)$ denotes the set of BSs which interfere with UE $j$ (note that $i(j)\in \mathcal{B}(j)$). The channel gain $h_{j,i(j)}$ is assumed to follow the Nakagami-m distribution \cite{1318888} with 
probability density 
\be
\label{Eq: Nakagam-m}
 f_H(h;\mu,\Omega)=\frac{2\mu^{\mu}}{\Gamma(\mu)\Omega^{\mu}}h^{2\mu-1}\textrm{exp}\left(-\frac{\mu}{\Omega}h^2\right),\;h\ge 0
\ee
where the parameters are $\mu=\frac{\mathbb{E}[h^2]^2}{\textrm{Var}(h^2)}$, $\Omega=\mathbb{E}[h^2]$ and $\Gamma(\cdot)$ is the Gamma function. 
Moreover, $\eta\ge 2$ is the path-loss factor. Let $N_0$ denote the random noise power spectrum density, then $\sigma^2=N_0W$ is the total noise power. $G_{j,i(j)}^{\rm UE}$ and $G_{j,i(j)}^{\rm BS}$ denote the UE and BS antenna gain between UE $j$ and BS $i(j)$ respectively. 
In this paper, we assume that both the BSs and UEs are equipped with directional antennas.  The antenna gain is modeled by a `keyhole' \emph{sectorized antenna model} with constant main-lobe gain $G^{\rm max}$ and side-lobe gain $G^{\rm min}$, i.e.,
\begin{align}
\label{Eq: sectorized antenna gain}
G(\theta) = \left\{\begin{array}{cc} G^{\rm max}, & |\theta| \leq \frac{\Delta \theta} {2} \vspace{0.15cm} \\
G^{\rm min}, & |\theta| > \frac{\Delta \theta} {2}
\end{array} \right.
\end{align} 
where $\Delta \theta$ is the  
beam width (in radian). 
 Moreover, each BS/UE antenna has a constant total power radiation gain of $E$,  i.e., $
\Delta\theta G^{\rm max} + \left(2\pi -\Delta\theta \right)G^{\rm  min}=E$.
WLOG, we set $E=1$. We further define the  \emph{main to side-lobe ratio (MSR)} of the antenna, denoted by $D$, as 
\be
\label{Eq: directivity}
D \eqdef\frac{G^{\rm max}}{G^{\rm min}}\,.
\ee
Given $D$ and $\Delta\theta $, the maximum and minimum antenna gain can be calculated as $G^{\rm min} = \left( (D-1)\Delta\theta   +2\pi   \right)^{-1}$ and $G^{\rm max} = DG^{\rm min}$. Usually, the MSR is measured in decibel, which is $ D\textrm{ (dB)}=10\lg D$. 
We assume that all the BSs have identical antenna gain parameters and all the UEs also have identical antenna gain parameters. Therefore, we use $G^{\rm BS,max},G^{\rm BS,min}$ and $ \Delta \theta^{\rm BS}$ to represent the BS antenna parameters and $G^{\rm UE,max},G^{\rm UE,min}$ and $ \Delta \theta^{\rm UE}$ to represent the UE antenna parameters. For ease of presentation, we define the \emph{equivalent channel gain} between UE $j$ and the serving BS $i(j)$ as
\be 
g_{j,i(j)}   \eqdef \frac{G_{j,i(j)}^{\rm UE}G_{j,i(j)}^{\rm BS}|h_{j,i(j)}|^2d_{j,i(j)}^{-\eta}   }{\sum_{\ell \in\mathcal{B}(j)\backslash\{i(j)\}}p_{j(\ell),\ell}G_{j,\ell}^{\rm UE}G_{j,\ell}^{\rm BS}|h_{j,\ell}|^2d_{j,\ell}^{-\eta}    +\sigma^2}
\ee
and then the SINR at UE $j$ can be conveniently written as $\textrm{SINR}_{j,i(j)}=g_{j,i(j)}p_{j,i(j)}$. The main notations used in this paper are summarized in Table \ref{table: notation} on the top of the next page.
\begin{table*}[t]
\caption{Summary of notations}
\centering
\small \begin{tabular}{|c ||c|} 
\hline 
\textrm{Notation} & \textrm{Description}\\
 \hline
$M;K$ & \textrm{total number of BSs; total number of UEs} \\
 \hline
$\mathcal{K}_i;K_i$ &  \textrm{set of UEs associated with BS $i$, $\mathcal{K}_i\subseteq [K],|\mathcal{K}_i|=K_i$}\\
\hline
$W;W_{\rm c}$ & \textrm{total bandwidth; center frequency}\\
\hline 
$j(i)$& \textrm{UE $j(i)$ selected/served by BS $i$, $j(i)\in \mathcal{K}_i$}\\
\hline 
$i(j)$& \textrm{BS $i(j)$ serving UE $j$, $j\in \mathcal{K}_i$}\\
\hline 
$p_{j(i),i};p_{j,i(j)}$& \textrm{transmit power of BS $i$ (or $i(j)$) to its selected UE $j(i)$(or $j$)}\\
\hline 
$\bar{p}_{j,i}$ & \textrm{average power consumption of UE $j$ (associated with BS $i$)}\\
\hline 
$p_i^{\rm max};p_i^{\rm avg}$ & \textrm{maximum/average power constraint of BS $i$}\\
\hline
$d_{j,i}; h_{j,i}$ & \textrm{distance/small-scale fading between BS $i$ and UE $j$}\\
\hline
$ g_{j,i}$ & \textrm{equivalent channel gain between  BS $i$ and UE $j$}\\
\hline 
$g_{j,i}^{\rm max}(k)$ & \textrm{maximum equivalent channel gain between BS $i$ and UE $j$ at epoch $k$}\\
\hline 
$ g_{j,i}^{\rm max}   $ & \textrm{maximum channel gain overall blocks and epochs}\\
\hline
$G_{j,i}^{\rm BS};G_{j,i}^{\rm UE}  $ & \textrm{BS/UE antenna gain between BS $i$ and UE $j$}\\
\hline 
$G^{\rm BS,max};G^{\rm BS, min}$ & \textrm{maximum (main-lobe)/minimum (side-lobe) BS antenna gain}\\
\hline 
$G^{\rm UE,max};G^{\rm UE, min}$ & \textrm{maximum/minimum UE antenna gain}\\
\hline 
$\Delta \theta^{\rm BS};\Delta \theta^{\rm UE} $ & \textrm{main-lobe width of BS/UE antenna}\\
\hline 
$\gamma_{j,i}(k);\bar{\gamma}_{j,i}$ & \textrm{auxiliary variables at epoch $k$; time averaged value of auxiliary variables}\\
\hline 
$Z_i(k);H_{j,i}(k)$ & \textrm{Virtual queue values at epoch $k$}\\
\hline 
$X_{j,i}(k,n);{X}_{j,i}(k)$ & \textrm{Throughput of UE $j$ at block $n$ of epoch $k$; throughput at epoch $k$ }\\
\hline 
$\bar{X}_{j,i}$ & \textrm{Time averaged throughput}\\
\hline 
$T_{j,i}^{\rm d}(k,n)$ & \textrm{Data transmission time of UE $j$ from BS $i$ at block $n$ of epoch $k$}\\
\hline
\end{tabular}
\label{table: notation}
\end{table*}

In the following, we focus on distributed beam scheduling schemes with power  adaptation, which means that each BS will optimize its own transmit power without the knowledge of the transmit powers of other BSs, i.e., there is no 
information exchange among different BSs. We assume that each BS and UE can only have one beam scheduled at a time so in each time slot, each BS can only transmit to at most one UE and each UE can only receive data from the associated BS.  Moreover, throughout this paper, all interference will be treated as additive Gaussian noise at the target UEs.

\subsection{Distributed Beam Scheduling \& Network Utility Maximization}
We consider a slotted system operating synchronously.  
We assume that each time {\em frame} (or {\em epoch}) consists of $N$ {\em blocks} and each block has $T^{\rm b}$ time {\em slots}. Therefore, each epoch has $T=NT^{\rm b}$ slots. We assume a block fading channel where the channel gains stay unchanged during each epoch and are i.i.d. over different epochs. Scheduling happens among different blocks of each epoch. The time-averaged expected 
throughput of UE $j$ from the corresponding serving BS $i(j)$ is given by
\be
\label{Eq: time averaged expected throughout}
\bar{X}_{j,i(j)}=\lim\limits_{t\to \infty} \frac{1}{t}\sum_{k=1}^t \mathbb{E}\left[X_{j,i(j)}(k)\right], 
\ee 
where the expectation is taken over the system randomness (e.g., fading channel, scheduling). $X_{j,i(j)}(k)$ is the throughput of UE $j$ from its associated  
BS $i(j)$ in epoch $k$ and is calculated as
\be 
\label{eq: throughput calculation each epoch}
X_{j,i(j)}(k)=\sum_{n=1}^NT^{\rm d}_{j,i(j)}(k,n)W\log\left(1+\textrm{SINR}_{j,i(j)}(k,n)\right),
\ee
where $T^{\rm d}_{j,i(j)}(k,n)$ denotes the data transmission time for UE $j$ during block $n$ of epoch $k$. 
In addition, $\textrm{SINR}_{j,i(j)}(k,n)$ represents the SINR at UE $j$ during block $n$ of epoch $k$. Since we have assumed that scheduling happens among different blocks, i.e., the selected UE and beam selection will stay unchanged during each block, the SINR of UE $j$, $\textrm{SINR}_{j,i(j)}(k,n)$ will stay unchanged during  block $n$ if the BS transmit powers do not change.

For the network utility, we adopt the $\alpha$-fairness utility model given by
\begin{align}
 \label{Eq: alpha faireness}
U_{\alpha}(x) \eqdef  \left\{\begin{array}{ll} \frac{x^{1-\alpha}}{1-\alpha}, \quad &\textrm{if }\alpha\ge 0,\alpha \ne 1, \vspace{0.15cm}\\
\log x, &\textrm{if } \alpha =1,
\end{array} \right.
\end{align} 
where $\alpha$ is a free parameter. In this paper, we use  $U(x) = \log x$ (with base $e$) as the utility function. It can be seen that $U(x)$ is a continuous, concave and strictly increasing  
function. The utility of each UE $j$, denoted by $u_j^{\rm UE}$, is defined as the logarithm of the time averaged expected throughout (See (\ref{Eq: time averaged expected throughout})) of that UE, i.e., 
$
u_j^{\rm UE}=U(\bar{X}_{j,i(j)}),\forall j\in[K].
$
The utility of each BS $i$, denoted by $u_i^{\rm BS}$, is defined as the sum utility of the UEs associated with that BS, i.e., 
$
u_i^{\rm BS}=\sum_{j\in\mathcal{K}_i}u_j^{\rm UE},\forall i\in[M].
$
The \emph{network utility} is then defined as the sum utility of all the BSs, i.e.,
\be 
\label{Eq: network utility}
\textrm{Network utility}\eqdef\sum_{i\in[M]}\sum_{j\in\mathcal{K}_i}U(\bar{X}_{j,i}).
\ee

Our goal is to design efficient distributed access strategies to maximize the network utility subject to peak and average power constraints of each BS. In particular, we aim to solve the following stochastic optimization problem with variables $p_{j,i}(k,n)$:
\begin{subequations}
\label{Eq: sum utility maximization}
\begin{align}
\max \quad & \sum_{i\in[M]}\sum_{j\in\mathcal{K}_i}U\left(\bar{X}_{j,i}\right)\label{Eq 1} \\
\textrm{s.t.} \quad  
& \sum_{j\in\mathcal{K}_i}\bar{p}_{j,i} \le Tp_i^{\rm avg},\;\forall i\in[M]\label{Eq 2}\\
& 0 \leq    \sum_{j\in\mathcal{K}_i} p_{j,i}(k,n)\le p^{\rm max}_i,\;\forall i\in[M],k\ge 1,n\in[N]\label{Eq 3}\\
  & a(k,n) \in\mathcal{A}(k,n),\; \forall k\ge 1 , \forall n\in[N] \label{Eq 4}
\end{align}
\end{subequations}
where $\bar{p}_{j,i} = \lim\limits_{t \rightarrow \infty}\frac{1}{t}\sum_{k = 1}^t \sum_{n=1}^N\mathbb{E}\left[T^{\rm d}_{j,i}(k,n)p_{j,i}(k,n)\right]$ represents the average power consumption of BS $i$ to UE $j$ at epoch $k$; $p_{j,i}(k,n)$ is  the transmit power from BS $i$ to UE $j$ at block $n$ of epoch $k$; 
$p_i^{\rm avg}$ and $p_i^{\rm max}$ represent the average and peak power constraints for BS $i$, respectively; $a(k,n)$ represents the instantaneous control action of the access strategy at block $n$ of epoch $k$ and $\mathcal{A}(k,n)$ is the action space which depends on the specific access strategy. 
We let $U^{\rm opt}$ denote the optimal value of the above optimization problem. We assume that the UE association is fixed, i.e., it has been determined by some exogenous mechanism prior to our design. Since we have assumed that each UE can connect to at most one BS at a time and each BS can transmit to at most one UE at a time, this excludes the use of Successive Interference Cancellation (SIC) techniques which may not be a common practice in real-world cellular systems.

\section{Proposed Approach}
\label{Section: Proposed Approach}
According to the Lyapunov optimization theory \cite{neely2010stochastic}, we transform the network utility maximization problem of (\ref{Eq: sum utility maximization}), which aims to optimize a sum of logarithm functions of the time averaged expected throughput of the UEs, into a new optimization problem (\ref{Eq: transformed sum utility maximization}) which  aims to optimize the time averaged expected logarithm function of the UE throughput. The purpose of doing this transformation is to apply the well-established Lyapunov drift-plus-penalty framework.  
Further, the transformed optimization problem can be solved via solving two sub-problems at each epoch together with the updating of the virtual queues to enforce BS power constraints.

\begin{figure*}[t]
\centering
\includegraphics[width=.95\textwidth]{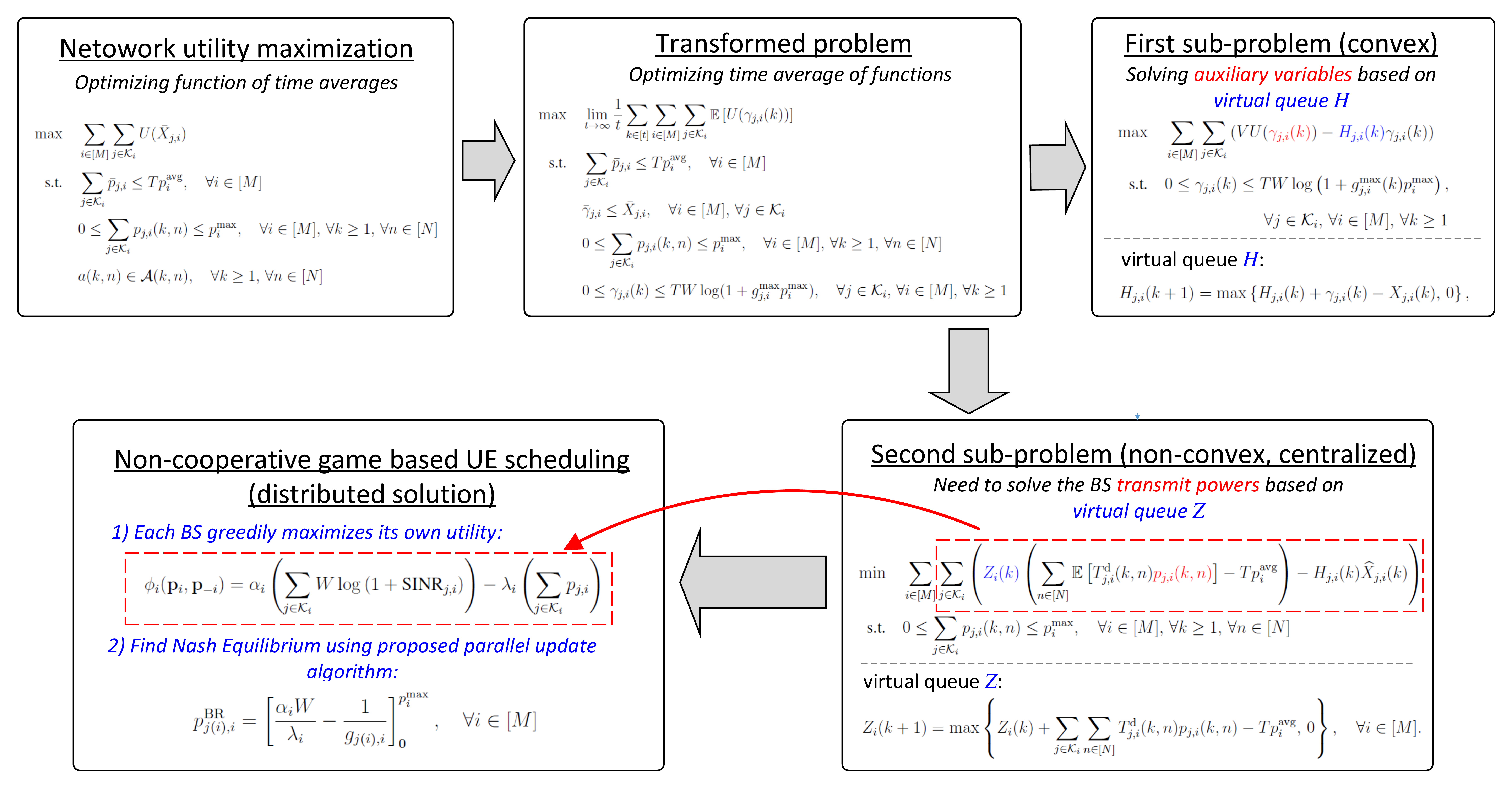}
\caption{\small Solution flow of the proposed distributed beam scheduling framework. The original network utility maximization problem which involves maximizing a function of the time averaged UE throughput is transformed into a new problem whose objective is to maximize the time averaged value of the utility function. The transformed problem is then decomposed into two sub-problems by introducing the auxiliary variables $\gamma_{j,i}(k)$ in each epoch. The first sub-problem is convex and thus can be easily solved. The second sub-problem is non-convex and aims to determine the BS transmit powers $p_{j,i}(k,n)$ in each block. We formulate the distributed beam scheduling problem as a non-cooperative game where each BS attempts selfishly to maximize its own payoff which is a function of the Lyapunov virtual queue parameters $Z_i(k),H_{j,i}(k)$ and the transmit powers of other BSs. We then propose a parallel update algorithm to find the Nash Equilibrium of the game which serves as a distributed solution to the second sub-problem. }
\label{fig_solution_flow}
\vspace{-0.8cm}
\end{figure*}

We formulate the distributed beam scheduling problem as a non-cooperative game and propose to solve the two sub-problems by solving  the Nash Equilibrium (NE). The payoff functions of the players (i.e., BSs) are determined by the objective functions of the two sub-problems and have a nice mathematical structure which guarantees the existence and uniqueness (under certain conditions) of the NE. A step-by-step relaxation from the original network utility maximization to the non-cooperative game-based distributed solution is given in   Fig.~\ref{fig_solution_flow}. 

\subsection{The General Lyapunov Optimization Framework}
\label{Subsection: The Lyapunov Transformation}
By introducing a set of $K$ auxiliary variables $\left\{\gamma_{j,i}(k): i\in[M], j\in \mathcal{K}_i \right\}$ in each epoch $k$, which represent the the average throughput of each UE in epoch $k$, the original optimization problem (\ref{Eq: sum utility maximization}) can be transformed into the following equivalent optimization problem with a time averaged objective function and variables $p_{j,i}(k,n)$ and $\gamma_{j,i}(k)$: 
\begin{subequations}
\label{Eq: transformed sum utility maximization}
\begin{align}
\max & \quad \lim_{t \to \infty} \frac{1}{t} \sum_{k\in[t]} \sum_{i\in[M]} \sum_{j\in\mathcal{K}_i} \EE\left[ U( \gamma_{j,i} (k))\right] \label{Eq: 11} \\
\text{s.t.} & \quad \sum_{j\in\mathcal{K}_i}  {\bar{p}_{j,i}} \leq  Tp_i^{\rm avg},\;\forall i\in[M]\label{Eq: 12} \\ 
& \quad \bar \gamma_{j,i} \leq \bar X_{j,i}, \; \forall i\in[M],\forall j\in\mathcal{K}_i \label{Eq: 13}\\
& \quad 0 \le  \sum_{j\in \mathcal{K}_i} p_{j,i}(k,n) \le p_i^{\rm max},\;\forall i\in[M], k\ge 1, n\in[N]\label{Eq: 14}\\
& \quad 0 \leq \gamma_{j,i}(k) \leq TW\log\left(1+ g_{j,i}^{\rm max}p_i^{\rm max}\right),\; \forall i\in[M] ,\forall j\in\mathcal{K}_i,k\ge 1 \label{Eq: 15}
\end{align}
\end{subequations}
where $g_{j,i}^{\rm max}$ denotes the maximum equivalent channel gain from BS $i$ to UE $j$ over all blocks and epochs, 
i.e., $g_{j,i}^{\rm max}\eqdef \max_{k,n} g_{j,i}(k,n)$. $\bar \gamma_{j,i}\eqdef\lim\limits_{t\to \infty}\frac{1}{t}\sum_{k=1}^t  \gamma_{j,i}(k)$ denotes the average value of the auxiliary variable $\gamma_{j,i}(k)$ over all epochs.

The above transformed optimization problem can be solved by solving two sub-problems at each epoch together with the updating of two \emph{virtual queues} to enforce the average and peak power constraints of the BSs \cite{neely2010stochastic}. In particular, we define two virtual queues $\{Z_i(k)\}_{k=1}^{\infty},\forall i\in[M]$ and $\{H_{j,i}(k)\}_{k=1}^{\infty}, \forall i\in[M],\forall j\in\mathcal{K}_i$ which are  updated at each epoch. The first virtual queue $\{Z_i(k)\}_{k=1}^{\infty}$ corresponds to the transmit powers $p_{j,i}(k,n)$ and is updated according to  
\begin{align}
\label{Eq: queue 1} 
Z_i(k+1) &=\max \Bigg\{Z_i(k) + \sum_{j\in\mathcal{K}_i}\sum_{n\in[N]}  T_{j,i}^{\rm d}(k,n)p_{j,i}(k,n) -Tp_i^{\rm avg},\,0 \Bigg\},\;\forall i\in[M].
\end{align}
The purpose of this virtual queue is to enforce the satisfaction of the average BS power consumption constraint (\ref{Eq: 12}). 
The second virtual queue $\{H_{j,i}(k)\}_{k=1}^{\infty}$ corresponds  to the auxiliary variables $\gamma_{j,i}(k)$ and is updated according to
\be
\label{Eq: queue 2} 
H_{j,i}(k+1) = \max\big\{ H_{j,i}(k)+\gamma_{j,i}(k)-X_{j,i}(k),\,0 \big\},  \;\forall i\in[M],\forall  j\in\mathcal{K}_i
\ee
which is used  to enforce the average constraint 
(\ref{Eq: 13}) on the auxiliary variables. With the definition of the virtual queues, we are now ready to present the two sub-problems.

The first sub-problem aims to solve the auxiliary variables $\gamma_{j,i}(k)$ in each epoch $k$:
\begin{subequations}
\label{Eq: sub-opt 1}
\begin{align}
\max \quad & \sum_{i\in[M]}\sum_{j\in\mathcal{K}_i} \big(V U({ \gamma_{j,i}(k)})  -{H_{j,i}(k)}\gamma_{j,i}(k)\big)\label{Eq: sub-opt 11} \\
\textrm{s.t.} \quad & 0\le \gamma_{j,i}(k)\le TW\log\left(1+g_{j,i}^{\rm max}(k)p_i^{\rm max}\right), \;\forall i\in[M], \forall j\in\mathcal{K}_i, k\ge 1\label{Eq: sub-opt 12} 
\end{align}
\end{subequations}
where $g_{j,i}^{\rm max}(k)$ denotes the maximum value of $g_{j,i}(k,n)$ in epoch $k$, i.e., $g_{j,i}^{\rm max}(k)   \eqdef\max_{n}g_{j,i}(k,n)$.\footnote{From the boundedness constraint (\ref{Eq: 15}), ideally, we want to upper bound $\gamma_{j,i}(k)$ by $\gamma_{j,i}(k)\le TW\log(1+g_{j,i}^{\rm max}p_i^{\rm max})$ instead of using $g_{j,i}^{\rm max}(k)$. However, for implementation, we want to solve the sub-problem (\ref{Eq: sub-opt 1}) at each epoch, so it is impossible to get knowledge of the equivalent gains in the future epochs. Therefore, we use $g_{j,i}^{\rm max}(k)$ as a substitute of $g_{j,i}^{\rm max}$. Furthermore, $g_{j,i}^{\rm max}(k)$ also needs to be estimated at the beginning of epoch $k$. We can adopt any large enough finite constant which is an upper bound on $g_{j,i}^{\rm max}(k)$ as a substitute of $g_{j,i}^{\rm max}(k)$. According to \cite{neely2010stochastic}, the effect of this estimate is negligible and the optimality of problem (\ref{Eq: transformed sum utility maximization}) will not be affected if the chosen constant is large enough. Note that there is no need for each BS to know the exact value of the equivalent channel gain $g_{j,i}(k,n)$ in (\ref{Eq: sub-opt 1}). In fact, it is impossible for BS $i$ to know $g_{j,i}(k,n)$ in a distributed system as the equivalent channel gain depends on the transmit powers of all interfering BSs.}
The parameter $V$ is a constant that can be tuned to find a desirable trade-off between optimality gap (to the original problem  (\ref{Eq: sum utility maximization})) and convergence speed. It can be seen that for fixed virtual queue status in epoch $k$, the sub-problem (\ref{Eq: sub-opt 1}) is a convex optimization problem. Moreover, the first sub-problem interacts with the virtual queue $\{H_{j,i}(k)\}_{k=1}^{\infty}$ as follows. 
From (\ref{Eq: sub-opt 11}), we see that if the queue status $H_{j,i}(k)$ is large at the current epoch $k$, which implies that the average value (up to the current epoch) of the auxiliary variable $\gamma_{j,i}$ is large, then maximizing the objective function (\ref{Eq: sub-opt 11}) will yield a small $\gamma_{j,i}(k)$ which reduces the average value of the auxiliary variables and enforces the satisfaction of the time averaged constraint  $\bar{\gamma}_{j,i}\le \bar{X}_{j,i}$ of (\ref{Eq: 13}).
 
The second sub-problem aims to solve the transmit powers $p_{j,i}(k,n)$ in each block $n$ of epoch  $k$:
\begin{subequations}
\label{Eq: sub-opt 2}
\begin{align}
\min  \quad & \sum_{i\in[M]}\sum_{j\in\mathcal{K}_i}   \left( \sum_{n\in[N]}     
\mathbb{E}\left[T_{j,i}^{\rm d}(k,n){p_{j,i}(k,n)}\right] -Tp_i^{\rm avg} \right) Z_{i}(k)
-H_{j,i}(k)\widehat{X}_{j,i}(k)
\label{Eq: sub-opt 21} \\
\textrm{s.t.} \quad & 0\le \sum_{j\in\mathcal{K}_i} p_{j,i}(k,n)\le p_i^{\rm max},\; \forall i\in[M],k\ge 1,\forall n\in [N] \label{Eq: sub-opt 22}
\end{align}
\end{subequations}
where $$\widehat{X}_{j,i}(k)\eqdef\sum_{n=1}^N\mathbb{E}\left[T_{j,i}^{\rm d}(k,n)W\log(1+ \textrm{SINR}_{j,i}(k,n))\right]$$ denotes the expected throughput achieved by UE $j$ (served by BS $i$) in epoch $k$ and $\textrm{SINR}_{j,i}(k,n)=g_{j,i}(k,n)p_{j,i}(k,n)$. This sub-problem interacts with the virtual queue $\{Z_i(k)\}_{k=1}^{\infty}$ as follows. 
From (\ref{Eq: sub-opt 21}), it can be seen that when the queue status $Z_i(k)$ is large in the current epoch $k$, implying the time averaged power consumption (up to the current epoch) of BS $i$ is high, then minimizing the objective function (\ref{Eq: sub-opt 21}) will yield some small values of power allocation to the UEs of BS $i$, which reduces the average power consumption of BS $i$ and therefore enforces the satisfaction of the average power constraint (\ref{Eq: 12}).

By solving the two sub-problems (\ref{Eq: sub-opt 1}) and (\ref{Eq: sub-opt 2}) in each epoch and updating the virtual queues using (\ref{Eq: queue 1}) and (\ref{Eq: queue 2}), the following proposition for the performance guarantee of this approach can be obtained straightforwardly \cite{neely2010stochastic}:
\begin{proposition}
\label{proposition: optimality gap -D/V}
\emph{
Let $\bar{X}_{j,i}^{\rm sub-opt}, \forall i\in[M],\forall j\in\mathcal{K}_i$ be the optimal average throughput achieved by solving the two sub-problems (\ref{Eq: sub-opt 1}), (\ref{Eq: sub-opt 2}) in each epoch. Given that the utility function $U(x)=\log x$ 
and the system state is i.i.d. over every epoch, then all the constraints in the transformed problem (\ref{Eq: transformed sum utility maximization})
can be satisfied and
\be
\label{Eq: optimality gap -D/V}
\sum_{i\in[M]}\sum_{j\in\mathcal{K}_i}U(\bar{X}_{j,i}^{\rm sub-opt})\ge U^{\rm opt} -\frac{B}{V},
\ee
where $U^{\rm opt}$ is the maximum utility of the original optimization problem (\ref{Eq: sum utility maximization}) and $B$ is some constant.
\hfill$\square$}
\end{proposition}

It can be seen from Proposition \ref{proposition: optimality gap -D/V} that if $V$ is large, then the proposed approach can achieve almost the same optimal network utility as the original problem. We observe that the first sub-problem (\ref{Eq: sub-opt 1}) is a convex optimization problem which can be easily solved distributedly.  
However, the second sub-problem (\ref{Eq: sub-opt 2}) is a stochastic non-convex optimization problem in general and it is required to solve this sub-problem 
distributedly among the BSs.
Hence, finding the optimal solution for (\ref{Eq: sub-opt 2}) is challenging. In what follows, we provide a non-cooperative game-based approach to solve the distributed scheduling problem. 
We next explain intuitively how the second  sub-problem (\ref{Eq: sub-opt 2}) is connected to non-cooperative games. When the virtual queue status $Z_i(k),H_{j,i}(k),\forall i\in[M],\forall j\in\mathcal{K}_i$ are given (this is because the status of the two virtual queues are determined by the data transmission of the previous epoch and is independent of the BS transmit powers at the current epoch), the objective function (\ref{Eq: sub-opt 21}) becomes minimizing the difference between the total power consumption and the average throughput weighted by the virtual queue status across all BSs. This is equivalent to maximizing the sum of a (\ref{Eq: general GT payoff function})-like (See Section \ref{Subsection: Non-cooperative game formulation}) payoff function for all BSs with pre-determined and optimal ``weights" $\alpha_i$ and $\lambda_i$ (called pricing factors). We
aim to solve this problem in a distributed manner, i.e., BSs do not coordinate in determining their transmit powers. 
Instead, each BS myopically maximizes its own payoff by choosing its transmit powers based on the measured interference from other BSs. This non-cooperative game theory provides a straightforward approximate solution to such a distributed optimization problem. 
In the following subsection, we will provide a detailed description of the non-cooperative game-based formulation.

\subsection{Non-cooperative Game-based Formulation}
\label{Subsection: Non-cooperative game formulation}
The distributed nature of the beaming scheduling task falls into the scope of the non-cooperative games in which a set of players tries to maximize their individual payoff based on the decisions of other players. In this section, we propose a distributed beam scheduling algorithm by formulating the scheduling problem as a non-cooperative game in which the BSs are the players each having a payoff function which is the aggregate throughput achieved by the associated UEs plus a power consumption penalty term. Each player then tries to maximize its own payoff based on the power allocation decisions and the CSI. 
This game happens in each scheduling unit, i.e., a block. By finding the Nash Equilibrium (NE) of the non-cooperative power allocation game, the proposed scheduling algorithm provides a good distributed approximation to the sub-problem (\ref{Eq: sub-opt 2}). In other words, the sub-problem (\ref{Eq: sub-opt 2}) fits naturally into the scope of non-cooperative games in game theory \cite{980097}, where instead of pre-defining the pricing factors $\alpha_i,\lambda_i$ as in most of the work in the literature, the pricing factors in this problem are determined by the status of the virtual queues. Before proceeding, we first describe the non-cooperative games in a more general sense, providing several key properties of the game and then adapt the game theory framework to our specific scheduling problem in each epoch.

Consider the power allocation game $\mathcal{G}=\left\langle [M],\{\mathcal{P}_i\}_{i\in[M]},\{\phi_i\}_{i\in[M]}\right\rangle $ in our considered network model described in Section \ref{Subsection: Network Model}, in which the set of $M$ BSs are the players. WLOG, 
we consider the case where each BS is associated with the same number of UEs, i.e., $K_i=K/M,\forall i\in[M]$. The action space of BS $i\in[M]$, denoted by $\mathcal{P}_i$, is defined as 
\be
\label{eq: action space}
\mathcal{P}_i\eqdef\left\{\pv_i: 0\le  \sum_{j\in\mathcal{K}_i} p_{j,i}\le   p_i^{\rm max}, p_{j,i}\ge 0,\forall j\in\mathcal{K}_i\right\},
\ee
where $\pv_i\eqdef(p_{j,i})_{j\in\mathcal{K}_i}\in \mathbb{R}_+^{K/M}$ 
denotes the power allocation profile for BS $i$, i.e., the power allocation to each UE associated with BS $i$. Let $\pv_{-i}\eqdef\left\{\pv_{i'}:i'\in[M]\backslash\{i\}\right\}$ denote the power profile for all BSs expect BS $i$. The payoff function $\phi_i$ of BS $i$ is defined as
\begin{align}
\label{Eq: general GT payoff function}
& \phi_i(\pv_i,\pv_{-i})= \alpha_i\Bigg(\sum_{j\in\mathcal{K}_i}W\log \left(1+\textrm{SINR}_{j,i}  \right)\Bigg)- \lambda_i\Bigg( \sum_{j\in\mathcal{K}_i}p_{j,i}\Bigg),
\end{align}
in which $\textrm{SINR}_{j,i}=g_{j,i}p_{j,i}$ is the received SINR at UE $j$ of BS $i$ and $\alpha_i\ge 0,\lambda_i\ge 0$ are some non-negative weights referred to as pricing factors. This payoff function has an intuitive interpretation that it aims to maximize the throughput while penalizing the over consumption of power which is consistent with the average power constraints of the BSs. In general, the pricing factors $\alpha_i$ and $\lambda_i$ can be tuned to find a desirable trade-off between throughput and power consumption. For example, in \cite{shi2018non} a scenario where multiple radar and communication systems coexist, a game theoretic method was proposed to control the interference where the pricing factors are adjusted heuristically according to the achieved SINR at the communication system. 
In our proposed distributed scheduling algorithm, however, the pricing factors  $\alpha_i,\lambda_i$ are updated according to the status of the virtual queues determined by (\ref{Eq: queue 1}), (\ref{Eq: queue 2}) and the first sub-problem (\ref{Eq: sub-opt 1}). Such a choice of the pricing factors ensures the constant-factor optimality of the proposed framework due to the Lyapunov transformation (See Lemma \ref{lemma: GT optimality gap}).
Next we formally present the definition of the Nash Equilibrium for the game $\mathcal{G}$ through the best response function.

\begin{defn}[\textbf{Best Response, BR}]
\emph{The Best Response for each BS $i$, denoted by $\pv_i^{\rm BR}$, given the power profiles $\pv_{-i}$ of all other BSs, is defined as a power profile of BS $i$ such that its payoff is maximized, i.e.,
$
\label{Eq: BR}
\phi_i\left(\pv_i^{\rm BR},\pv_{-i}\right) \ge \phi_{i}\left( \pv_i,\pv_{-i} \right),\forall \pv_i\in\mathcal{P}_i.
$
Moreover, the Best Response function for BS $i$, as a function of the power profiles $\mathbf{p}_{-i}$, is defined as 
$
\label{Eq: BRF}
\pv_i^{\rm BR}(\pv_{-i})={\rm arg}\max_{\pv_i\in \mathcal{P}_i}\phi_i\left( \pv_i, \pv_{-i} \right).
$
}
\end{defn}

With the definition of BR, the Nash Equilibrium of $\Gc$ is then defined as follows.
\begin{defn}[\textbf{Nash Equilibrium, NE}]
\emph{The Nash Equilibrium of the distributed scheduling game $\mathcal{G}$ is defined as a power allocation profile $\{\pv_i^{\star}\}_{i\in[M]}$ such that each BS's power allocation profile is the Best Response to the power allocations of all other BSs, i.e., $\forall i\in[M]$:
\be
\phi_i(\pv_i^{\star},\pv^{\star}_{-i}) \ge  \phi_i(\pv_i,\pv^{\star}_{-i}),\quad \forall \pv_i\in\mathcal{P}_i
\ee 
}\end{defn} 
From the above definition, we see that NE is a power allocation for which no BS has the incentive to unilaterally deviate from it to obtain a better individual payoff. Solving the NE for the non-cooperative game $\mathcal{G}$ is essentially solving a set of $M$ coupled optimization problems where the objective function for each of these optimization problems is the payoff of the  corresponding BS which depends also on the power allocation of other BSs.

\subsection{Existence and Uniqueness of Nash Equilibrium}
\label{Subsection: Existence and Uniqueness of Nash Equilibrium}
In this section we discuss the properties of the NE of the power allocation game $\mathcal{G}$ defined in Section \ref{Subsection: Non-cooperative game formulation}. More specifically, given the structure of the game, we prove that $\mathcal{G}$ always admits at least one NE for arbitrary channel realizations.
We further provide sufficient conditions which guarantee the uniqueness of the NE by establishing an equivalence between the non-cooperative game and a corresponding Variational Inequality (VI) problem \cite{facchinei201012}. Borrowing existing results on the uniqueness of solutions of the VI problem, we are able to prove the uniqueness of NE.

Since we have assumed no use of SIC techniques, 
each BS can only transmit to at most one UE during a block in the proposed distributed scheduling algorithm. To choose which UE to serve,
multiple
approaches such as random selection and Round Robin can be used. However, multiple BSs can transmit to their designated UEs simultaneously. In this case, the aggregate interference  from other transmitting BSs will be simply treated as Gaussian noise. Under this scheduling model, the BR function for each BS is given in Lemma \ref{Lemma: BR}. Recall that for any BS $i$, we let $j(i)$ denote the UE which is served by this BS; For any UE $j$, we use $i(j)$ to denote the BS which is responsible to serve this UE.
\begin{lemma}
\label{Lemma: BR}
\emph{
Suppose that at most one UE can be served by each BS at any time, given the payoff function defined in  (\ref{Eq: general GT payoff function}), the Best Response of BS $i$, $\pv_i^{\rm BR}\eqdef\left(p_{j,i}^{\rm BR} \right)_{j\in\mathcal{K}_i}$, is given by
\be
\label{Eq: Water-filling BR}
p_{j(i),i}^{\rm BR}=\left[\frac{\alpha_iW}{\lambda_i}-\frac{1}{g_{j(i),i}} \right]_0^{p_i^{\rm max}},\;\forall i\in[M]
\ee
where UE $j(i)$ is the only UE served by BS $i$. We have $p_{j',i}^{\rm BR}=0,\forall j'\in \mathcal{K}_i\backslash\{ j(i)\}$, and  $g_{j(i),i}=\frac{G^{\rm UE}_{j(i),i}G^{\rm BS}_{j(i),i}|h_{j(i),i}|^2d_{j(i),i}^{-\eta}}{\sum_{\ell\in[M]\backslash \{i\}}G^{\rm UE}_{j(i),\ell}G^{\rm BS}_{j(i),\ell}|h_{j(i),\ell}|^2d_{j(i),\ell}^{-\eta}p_{j(\ell),\ell} +\sigma^2 }$ is the equivalent channel gain from BS $i$ to UE $j(i)$. \hfill $\square$
}\end{lemma}
\begin{IEEEproof}
See Appendix \ref{Appendix: Proof of Lemma 1}.
\end{IEEEproof}

Based on the Best Response function derived in the above lemma, solving the NE can be formulated as solving a fixed point equation. In particular, if the NE of $\mathcal{G}$ exists, then it must satisfy a set of non-linear equations specified by (\ref{Eq: Water-filling BR}). It can be seen that the NE $\{\mathbf{p}^{\star}_i\}_{i\in[M]}$ is a fixed point of the Euclidean projection mapping defined by (\ref{Eq: Water-filling BR}). Therefore, the NE can be found  using the fixed point iteration algorithm \cite{rhoades1976comments}. In our scheduling algorithm, BR based iteration method can be used to find the NE based on the interaction (via interference) among different BSs. We next prove the existence and uniqueness of the NE of the  considered game.

\begin{lemma}[\textbf{Existence of NE}] 
\label{Lemma: existence of NE}
\emph{Based on the considered scheduling model, the game $\mathcal{G}=\left\langle [M],\{\mathcal{P}_i \}_{i\in[M]},\{\phi_i\}_{i\in[M]}\right\rangle $ always admits at least one pure strategy NE\footnote{A pure strategy NE is a NE in which each BS chooses a certain power allocation profile with probability one.} for any  $\alpha_i,\lambda_i\ge 0,\forall i\in[M]$ and any set of channel realizations. \hfill $\square$
}\end{lemma} 
\begin{IEEEproof}
See Appendix \ref{appendix: proof of lemma 2}.
\end{IEEEproof}

Since the NE of $\mathcal{G}$ always exists, we are interested in finding a set of sufficient conditions under which the equilibrium is unique. The uniqueness of NE is established via the connection to the Variational Inequality (VI) problem \cite{facchinei201012}. Before proceeding to prove the uniqueness of the NE, we give a brief description of the VI problem. Given a closed and convex set $\mathcal{M}\subseteq \mathbb{R}^n$ and a mapping $\mathbf{F}:\mathcal{M} \mapsto \mathbb{R}^n$, the VI problem, denoted by $\textrm{VI}(\mathcal{M},\mathbf{F})$, aims to find a vector $\xv^{\star}\in\mathcal{M}$ such that 
$ (\yv -\xv^{\star})^{\sf T}\mathbf{F}(\xv^{\star})\ge 0,\forall \yv\in \mathcal{M}$, 
in which $\xv^{\star}$ is called the solution of $\textrm{VI}(\mathcal{M},\mathbf{F})$. For our considered non-cooperative game $\mathcal{G}$, the corresponding VI problem can found as follows. Let $\mathcal{P}\eqdef\prod_{i=1}^M\mathcal{P}_i$ denote the product space. Recall that $j(i)$ is the UE selected by BS $i$ to transmit to. Let $\nu(i) \eqdef {\sf mod}\left(j(i),K/M \right) $ be the index of UE $j(i)$ among the UEs associated with BS $i$. We define a vector function $\mathbf{F}:\mathcal{P} \mapsto \mathbb{R}^{1\times M}$ as $\mathbf{F}(\pv)\eqdef \left[\mathbf{F}_1(\mathbf{\pv}),\mathbf{F}_2(\mathbf{\pv}),\cdots,\mathbf{F}_M(\mathbf{\pv}) \right]\in \mathbb{R}^{\frac{K}{M}\times M}$ in which $\mathbf{F}_i(\pv),\forall i\in[M]$ is defined as 
\begin{subequations}
\label{Eq: VI mapping}
\begin{align}
\mathbf{F}_i(\pv)&\eqdef-\nabla_{\pv_i}\phi_i(\pv_i,\pv_{-i})\\
&= \left[ \mathbf{0}^{\nu(i)-1}, -\frac{\partial \phi_i(\pv_{i},\pv_{-i})}{\partial p_{j(i),i}},\mathbf{0}^{\frac{K}{M}-\nu(i)}   \right]^{\sf T}\\
&= \left[ \mathbf{0}^{\nu(i)-1}, \lambda_i-\frac{\alpha_ig_{j(i),i}W}{1+g_{j(i),i}p_{j(i),i} },\mathbf{0}^{\frac{K}{M}-\nu(i)}   \right]^{\sf T},
\end{align}
\end{subequations}
i.e., the only non-zero entry in the $\nu(i)^{\rm th}$ position of $\mathbf{F}_i(\pv)$ represents the first-order derivative of the payoff function $\phi_i$ w.r.t. the transmit power of BS $i$ to the selected UE $j(i)$. 
Note that the selection of which UE to serve by each BS is determined by some exogenous mechanism and here we just assume that the UE selection is fixed, i.e., each BS $i$ selects UE $j(i)$. 
It was shown in \cite{facchinei2007finite} that the game $\mathcal{G}$ is equivalent to the VI problem $\textrm{VI}(\mathcal{P},\mathbf{F})$. A direct consequence of this equivalence is that  if the mapping  $\mathbf{F}$ is a \emph{uniformly P-function}, then VI$(\mathcal{P},\mathbf{F})$ has a unique solution, which implies that the game $\mathcal{G}$ admits a unique equilibrium. This result is formally described in Proposition \ref{Proposition: unique solution of VI(P,F)}. In the following, we introduce two definitions which are useful in proving the uniqueness of the equilibrium.
 
\begin{defn}[\textbf{Uniformly P-function}]
\emph{The mapping $\mathbf{F}$ is said to be a uniformly P-function on $\mathcal{P}$ if there exists a constant $C^{\rm up}>0$ such that for any two power allocation profiles $\pv=(\pv_i)_{i=1}^M\in\mathbb{R}_+^{\frac{K}{M}\times M}$ and $\pv'=(\pv_i')_{i=1}^M\in\mathbb{R}_+^{\frac{K}{M}\times M}$, it holds that
 \be
 \max_{1\le i\le M} \left(\pv_i -\pv_i' \right)^{\sf T}\left(\mathbf{F}_i(\pv)-\mathbf{F}_i(\pv ') \right)\ge C^{\rm up}\| \pv-\pv' \|_2^2.
 \ee
in which $\| \pv-\pv' \|_2$ represents the Frobenius norm of the matrix $\pv-\pv'$.}\end{defn}

\begin{defn}[\textbf{P-matrix}]\emph{
A matrix $\mathbf{A}\in \mathbb{R}^{n\times n}$ is called a P-matrix if every principal minor of $\mathbf{A}$ is positive.}
\end{defn}

\begin{proposition}[\textbf{Uniqueness of Solution to VI$(\mathcal{P},\mathbf{F})$}, \cite{facchinei2007finite}] 
\label{Proposition: unique solution of VI(P,F)}
\emph{If each $\mathcal{P}_i,\forall i\in[M]$ is a closed convex set and $\mathbf{F}$ is a continuous uniformly P-function on $\mathcal{P}$,then $\textrm{VI}(\mathcal{P},\mathbf{F})$ has a unique solution. Equivalently, the game $\mathcal{G}$ admits a unique NE. \hfill $\square$}
\end{proposition}

Next we introduce the matrix $\mathbf{Q}\eqdef[Q_{p,q}]\in\mathbb{R}^{M\times M}$ which is useful in studying the sufficient conditions guaranteeing the uniqueness of the equilibrium. $\mathbf{Q}$ is defined as
\begin{align}
\label{Eq: Q matrix}
& {Q}_{p,q}= \left\{\begin{array}{ll} \alpha_p W, & \textrm{if}\; p=q \\
 -\alpha_p W\left|\frac{\hbar_{j(p),q}}{\hbar_{j(q),q}}\right|^2\left(   1+ \frac{ \sum_{i\in[M]}|\hbar_{j(q),i}|^2p_i^{\rm max} }{\sigma^2} \right), & \textrm{if}\; p\ne q
\end{array} \right.
\end{align} 
where $\hbar_{j,i}\eqdef \sqrt{G^{\rm UE}_{j,i}G^{\rm BS}_{j,i}|h_{j,i}|^2d_{j,i}^{-\eta}}$. For a unified notation, we further denote $\widehat{\hbar}_{j(p),q}\eqdef\frac{\hbar_{j(p),q}}{\hbar_{j(p),p}}$. Note that $\widehat{\hbar}_{j(p),p}=1,\forall p\in[M]$. With such a specification of $\mathbf{Q}$, we are ready to present the uniqueness results in the following Theorem.

\begin{theorem}[\textbf{Sufficient Conditions on the Uniqueness of NE}]
\label{Theorem: Sufficient Conditions on the Uniqueness of NE}
\emph{ If the matrix $\mathbf{Q}$ defined in (\ref{Eq: Q matrix}) is a P-matrix, then the mapping $\mathbf{F}$ is a uniformly P-function. Consequently, the game $\mathcal{G}$ admits a unique NE. \hfill $\square$
}\end{theorem}
\begin{IEEEproof}
See Appendix \ref{Appendix: proof of Thm sufficient conditions}.
\end{IEEEproof}
\begin{remark}
Theorem \ref{Theorem: Sufficient Conditions on the Uniqueness of NE} gives a sufficient condition which guarantees the existence and uniqueness 
of the NE of the game $\mathcal{G}$. Since the matrix $\mathbf{Q}$ only depends on the parameters $\alpha_i,i\in[M]$ and the channel realization, it is possible that $\mathbf{Q}$ is a P-matrix. For example, due to structure of $\mathbf{Q}$ where all diagonal elements are equal to the constant $\alpha_pW$ while all off-diagonal elements are negative numbers depending on the channel gains, we notice that if all the channel gains are small enough, 
every principal minor of $\mathbf{Q}$ will be positive, making $\mathbf{Q}$ a P-matrix. 
\end{remark}

\subsection{Non-cooperative Game Based Beam Scheduling}
\label{Subsection: Non-cooperative Game Based Beam Scheduling}
With the general non-cooperative game-based formulation in Section \ref{Subsection: Non-cooperative game formulation}, we are
ready to present the proposed distributed beam scheduling algorithm. Recall that beam scheduling happens in each block of an epoch. To maximize the network utility, we aim to solve the two sub-problems (\ref{Eq: sub-opt 1}) and (\ref{Eq: sub-opt 2}) in a distributed manner at the beginning of each epoch. Recall that the first sub-problem is convex and can be  
solved by letting each BS perform an independent optimization of its own utility. The proposed distributed scheduling algorithm for solving sub-problem (\ref{Eq: sub-opt 2}) 
is as follows. At the beginning of each epoch, each BS $i\in[M]$ uniformly selects one UE $j(i)\in\mathcal{K}_i$ at random to transmit to until the end of the current epoch.
Therefore, the peak power constraint of (\ref{Eq: 14})  can be simplified as $0\le p_{j(i),i}\le p_i^{\rm max},\forall i\in[M]$. After the UE selection, the beams are aligned for BS $i$ and UE $j(i)$ if the transmit power is not zero.\footnote{If the transmit power of BS $i$ equals zero, which is possible in the game-based power update algorithm, then there is no need for beam generation between BS $i$ and UE $j(i)$.} In particular, BS $i$ aligns its beam with UE $j(i)$, i.e., UE $j(i)$ will lie in the center of the BS beam. UE $j(i)$ also aligns its beam with BS $i$.
All BSs will transmit to their designated UEs at the same time using the same spectrum. Therefore, BSs 
 interfere with each other. 
We assume that all BSs are synchronized which can be achieved by aligning timing with GPS. Since BSs are transmitting to their individually selected UEs throughout the entire epoch, for BS $i$, the data transmission time is $T^{\rm d}_{j(i),i}(k,n)=T^{\rm b}$ and $T_{j',i}^{\rm d}(k,n)=0,\forall j'\in\mathcal{K}_i\backslash\{j(i)\},\forall n\in[N]$. As a result, the objective function of the second sub-problem (\ref{Eq: sub-opt 2}) becomes\footnote{{Here we omitted the term $-\sum_{j\in\mathcal{K}_i}Z_i(k)Tp_i^{\rm avg}=-{KTZ_i(k)p_i^{\rm avg}}/{M}$ which is a constant. Therefore, removing this term from the objective function does not affect the solutions of the optimization problem.}}
\begin{subequations}
\label{Eq: sub-opt 2 with GT specialization, maximization}
\begin{align}
\max \quad & \sum_{i\in[M]}\sum_{n\in[N]}  H_{j(i),i}(k)T^{\rm b}W\log\left(1+\textrm{SINR}_{j(i),i}(k,n)  \right) -Z_i(k)T^{\rm b}p_{j(i),i}(k,n)   \\
\textrm{s.t.} \quad & 0\le p_{j(i),i}(k,n)\le p_i^{\rm max},\; \forall i\in[M],k\ge 1,\forall n\in[N].
\end{align}
\end{subequations}
We now solve the optimization problem (\ref{Eq: sub-opt 2 with GT specialization, maximization}) in each block distributedly using the proposed  game-based approach presented in Sections~\ref{Subsection: Non-cooperative game formulation} and~\ref{Subsection: Existence and Uniqueness of Nash Equilibrium}. In particular, in each block $n$ of epoch $k$, each BS $i$ aims to maximize the following payoff  function:
\begin{align}
\label{Eq: Payoff function specilization} 
& \phi_i \left({\pv_i(k,n),\pv_{-i}(k,n)}\right)= \alpha_iW\log\left(1+\textrm{SINR}_{j(i),i}(k,n)  \right) -\lambda_i p_{j(i),i}(k,n),
\end{align}
with 
\be 
\alpha_i \eqdef H_{j(i),i}(k)T^{\rm b},\quad  \lambda_i\eqdef Z_i(k)T^{\rm b}
\ee
where $\pv_i(k,n)$ is the power allocation profile for BS $i$. It can be seen that this payoff function fits exactly in the non-cooperative game  formulation (\ref{Eq: general GT payoff function}) with pricing factors  $\alpha_i = H_{j(i),i}(k)T^{\rm b}$ and $\lambda_i=Z_i(k)T^{\rm b}$. Let $\mathcal{G}{(k,n)}$ denote the power allocation game whose payoff function is defined by (\ref{Eq: Payoff function specilization}) and the {action space} for each BS $i$ is defined as
\begin{align}
\label{eq: GT BS i action space}
\mathcal{P}_i  &\eqdef\big\{\pv_i(k,n)\eqdef\big(p_{j,i}(k,n)\big)_{j\in\mathcal{K}_i}: 0\le p_{j,i}(k,n)\le p_i^{\rm max}, \forall i\in[M],\forall j\in\mathcal{K}_i \big\}. 
\end{align}
Each BS $i$ also maintains the virtual queues $\{Z_i(k)\}_{k=0}^{\infty}$ and $\{H_{j,i}\}_{k=0}^{\infty},\forall j\in\mathcal{K}_i$ in order to perform the   distributed scheduling. 

The Nash Equilibrium of the game $\mathcal{G}(k,n)$ can be found by performing a  standard parallel updating algorithm (See Algorithm \ref{algorithm:1}) based on the interactions via interference among different BSs \cite{scutari2008asynchronous}\footnote{{Other than the parallel updating algorithm, sequential updating in which the BSs update their transmit powers one after another in a sequential way can also be used to find the NE. The difference mainly lies in the convergence speed.}}. In particular, within each block $n$, each BS $i$ adapts its transmit power to the designated UE $j(i)$ slot-by-slot based on the interference (plus noise) measured at UE $j(i)$ from all other interfering BSs. In this case, the SINR of each designated UE, i.e., $\textrm{SINR}_{j(i),i}(k,n)$, will change from slot to slot. Therefore, the throughput of each UE needs to be calculated in a slot-by-slot manner instead of calculated from block to block as in  (\ref{eq: throughput calculation each epoch}). The parallel updating algorithm is formally described in Algorithm \ref{algorithm:1}. For ease of notation, we ignore the epoch and block indices $(k,n)$ on the power allocation profiles and denote $\hbar_{j,i}\eqdef \sqrt{G^{\rm UE}_{j,i}G^{\rm BS}_{j,i}|h_{j,i}|^2d_{j,i}^{-\eta}},\forall i\in[M],\forall j\in[K]$. Algorithm \ref{algorithm:1} works as follows. At the beginning of each epoch, each BS randomly picks an initial power $p_{j(i),i}^{(0)}$ from the action space defined by (\ref{eq: GT BS i action space}) to transmit to its designated UE $j(i)$. In each slot $s$, UE $j(i),\forall i\in[M]$ measures the received interference plus noise $I_{j(i)}^{(s)}$ and then sends $I_{j(i)}^{(s)}$ and  the estimated direct channel gain $\hbar_{j(i),i}$ to BS $i$ through some feedback mechanism\footnote{Since we have assumed a block fading channel model where the channel gains do not change during each epoch, the direct channel gain $\hbar_{j,i}$ only needs to be feedback to BS $i$ once per epoch, which can be done by assigning a number of designated time slots at the beginning of each epoch. The measured interference $I_{j(i)}^{(s)}$ needs to be feedback to BS $i$ periodically in order to perform the power updates, which means that certain slots also need to be allocated periodically for feedback. To increase the downlink data efficiency, the duration between two consecutive BS power updates can be increased so that a larger portion of time can be devoted to data transmission instead of feedback.}. 
BS $i$ then calculates the equivalent channel gain $g_{j(i),i}^{(s)}= {|\hbar_{j(i),i}|^2}/{I_{j(i)}^{(s)}} $ and adapts its transmit power to $p_{j(i),i}^{(s+1)}$ for the next slot $s+1$ according to equation (\ref{eq: update formula in Algorithm}). Each BS repeats this process until the stop criterion is met.
The stop criterion of the updating algorithm is that if either two consecutive power profiles are very close to each other, i.e., a difference of $\sqrt{\epsilon}$ for some pre-defined threshold $\epsilon>0$ in Frobenius norm, or the number of iterations reaches the maximum, i.e., the number of time slots per block. If the algorithm stopped before the iteration index $s$ reaches its maximum value  $T^{\rm b}$, the transmit powers of the BSs will be equal to the output of the algorithm for the remaining time slots. Note that the parallel updating algorithm is performed at each block, therefore the output of the algorithm at the current block will serve as the initial input to the algorithm at the next block. To perform the distributed scheduling algorithm, BS $i$ needs to know the virtual queue status $Z_i(k)$ and $H_{j(i),i}(k)$, the measured interference plus noise $I_{j(i)}^{(s)}$ at UE $j(i)$ and the channel gain $h_{j(i),i}$. The channel gain $h_{j(i),i}$ can be estimated by sending some pilots to UE $j(i)$ and then fed back to BS $i$. Similarly, the measured interference $I_{j(i)}^{(s)}$  at UE $j(i)$ can be fed back to BS $i$.\footnote{{The system overhead due to the feedback of the channel gain and measured interference (plus noise) from the UEs is negligible since is does not scale with the downlink data transmission.}} In addition, because the virtual queues are maintained separately by each BS, all the above information is available to BS $i$. Since the direct channel gain $\hbar_{j(i),i}$ can be obtained by BS $i$ via the feedback mechanism, the equivalent channel gain $g_{j(i),i}^{(s)}$ can be calculated as $g_{j(i),i}^{(s)}=|\hbar_{j(i),i}|^2/I_{j(i),i}^{(s)}$ as used in (\ref{eq: update formula in Algorithm}).
\begin{algorithm*}
\setstretch{1.2}  
  \caption{Parallel Updating Algorithm}\label{algorithm:1}
  \begin{algorithmic}[1]
  \item \textbf{\textit{Input}}: Randomly pick a feasible initial  point $\pv^{(0)}\eqdef\left\{\pv_i^{(0)}\right\}_{i\in[M]}\in \mathcal{P}$. Set time slot index $s=0$.
  \item \textbf{\textit{Step 1:}} If $\|\pv^{(s+1)}-\pv^{(s)}\|_2^2\le \epsilon$ or $s\ge T^{\rm b}$, then \textit{Stop}. 
  \item \textbf{\textit{Step 2:}} Each BS $i\in[M]$ compute (simultaneously):
  \be
  \label{eq: update formula in Algorithm}
   p_{j(i),i}^{(s+1)}= \left[\frac{H_{j(i),i}(k)W}{Z_i(k)}-\frac{1}{g_{j(i),i}^{(s)}}     \right]_0^{p_i^{\rm max}},
  \ee  
  where $g_{j(i),i}^{(s)}= {|\hbar_{j(i),i}|^2}/{I_{j(i)}^{(s)}} $ is the equivalent channel between BS $i$ and UE $j(i)$ at time slot $s$ and $I_{j(i)}^{(s)}$ denotes the interference plus noise measured at UE $j(i)$ at slot $s$.
  \item   \textbf{\textit{Step 3:}} Set $s\leftarrow s+1$. Go back to \textbf{\textit{Step 1}}.
  \item \textbf{\textit{Output:}} Output $\pv^{(s)}$.
  \end{algorithmic}
\end{algorithm*}
The parallel updating algorithm is proved to converge under the same condition that guarantees the uniqueness of NE of $\mathcal{G}(k,n)$ by \cite{pang2010design} (See Proposition \ref{Proposition: convergence of parallel updating algorithm}). In fact, our simulation results showed that the proposed parallel updating algorithm converges very fast in general (in dozens of slots).

\begin{proposition}[\textbf{Proof of Convergence}, \cite{pang2010design}] 
\label{Proposition: convergence of parallel updating algorithm} 
\emph{The sequence $\{\pv^{(s)}\}_{s=0}^{\infty}$ generated by Algorithm \ref{algorithm:1} always converges. Furthermore, if the matrix $\mathbf{Q}$ defined in (\ref{Eq: Q matrix}) is a P-matrix, then the sequence $\{\pv^{(s)}\}_{s=0}^{\infty}$ converges to the unique NE of the game $\mathcal{G}(k,n)$.}
\end{proposition}

\subsection{Optimality Gap Analysis}
In this section we identify one important property of the proposed game-based scheduling algorithm and analyze its optimality gap to the optimal value of the original network utility maximization problem.

Let $U^{\rm game}(k)$ and $U^{\rm ideal}(k)$ denote the network utility achieved by the proposed scheduling algorithm and the ideal case (See Section~\ref{Section: ideal case}) respectively, in  epoch $k$. The following lemma states the optimality gap of the proposed scheduling algorithm to the original utility maximization problem.
\begin{lemma} (\textbf{Optimality Gap}) 
\label{lemma: GT optimality gap}
\textit{
Suppose that there is an additive gap $C\ge 0$ in utility between the proposed game-based approach and the ideal case at each epoch, i.e.,  $U^{\rm game}(k)\ge U^{\rm ideal}(k) - C,\forall k\ge 1$. Then 
\be 
\label{eq: GT opt gap}
\sum_{i\in[M]}\sum_{j\in\mathcal{K}_i}U(\bar X_{j,i}^{\rm game}) \ge U^{\rm opt} -\frac{B+C}{V},
\ee
where $\bar X_{j,i}^{\rm game}$ is the average throughput achieved by UE $j$ (of BS $i$) in the proposed scheduling algorithm, $U^{\rm opt}$ is the optimal value of the original problem (\ref{Eq: sum utility maximization}) and $B$ is some constant. \hfill $\square$
}
\end{lemma}
\begin{IEEEproof} 
See Appendix \ref{appendix: proof of lemma 3}.
\end{IEEEproof}

When multiple NE exist, since it is unknown which one of the proposed parallel update algorithm will converge to, so we choose $C$  to be the upper bound on the optimality gap among all possible equilibrium power allocations. See Appendix \ref{appendix: gap C analysis}  for a detailed discussion of the optimality gap.

\section{Numerical Evaluation}
\label{Section: Numerical Evaluation}
\subsection{Description of the Baseline Schemes}
One advantage of the proposed Lyapunov optimization framework is that it can admit a number of underlying MAC layer protocols including $p$-persistent protocol and the 802.11 CSMA/CA protocol. In the following, we will consider the algorithms designed based on these two underlying MAC protocols as the baseline schemes in order to show the performance gain of the proposed game-based algorithm. We also consider  an `ideal case' where we assume there is no interference among BSs. This ideal case provides a natural upper bound on the performance of the proposed and baseline schemes. Moreover, we consider two more baselines which are $p$-persistent and CSMA/CA protocol with random power allocations. In the following, we will first describe how these two protocols fit into the proposed Lyapunov stochastic optimization framework and then we present the numerical results.

\subsubsection{$p$-persistent Access Strategy} In this case, we solve the network utility maximization problem (\ref{Eq: sum utility maximization}) under the $p$-persistent access strategy. In particular, we solve the two sub-problems (\ref{Eq: sub-opt 1}) and (\ref{Eq: sub-opt 2}) together with the updating of the two virtual queues at the beginning of each epoch. The first sub-problem  (\ref{Eq: sub-opt 1}) is a convex optimization problem and can be efficiently solved using the MATLAB optimization toolbox CVX. The second sub-problem involves the random data transmission time $\mathbb{E}[T^{\rm d}_{j,i}(k,n)]$ which has to be determined by some underlying access strategy and has to be estimated at the beginning of each epoch. Based on an estimate of $\mathbb{E}[T^{\rm d}_{j,i}(k,n)]$ which is denoted by $\widetilde{T}^{\rm d}_{j,i}(k,n),\forall j\in\mathcal{K}_i,\forall n\in[N]$, each BS $i$ needs to independently minimize\footnote{{Note that once the estimated data transmission time $\mathbb{E}[T_{j,i}^{\rm d}(k,n)]$ are given, the joint optimization problem of (\ref{Eq: sub-opt 2}) is equivalent to the independent optimization of (\ref{Eq: pp BS indep optimization}) performed by each BS. This is because in the $p$-persistent protocol, only one BS is allowed to transmit at any given time and the power constraints are independent for each BS. A similar situation holds when solving the auxiliary variables $\gamma_{j,i}(k)$ from the first sub-problem (\ref{Eq: sub-opt 1}).}}
\be 
\label{Eq: pp BS indep optimization}
Z_{i}(k)    \left( \sum_{n\in[N]}     
\widetilde{T}^{\rm d}_{j,i}(k,n)p_{j,i}(k,n) -Tp_i^{\rm avg} \right) -H_{j,i}(k)\widehat{X}_{j,i}(k),
\ee
subject to the BS peak transmit power constraints $p_{j,i}(k,n)\le p_i^{\rm max},\forall j\in\mathcal{K}_i,\forall n\in[N]$. We have $\widehat{X}_{j,i}(k)=\sum_{n\in[N]}\widetilde{T}^{\rm d}_{j,i}(k,n)W\log(1+ \textrm{SNR}_{j,i}(k,n))$ and $\textrm{SNR}_{j,i}=g_{j,i}p_{j,i}(k,n)$ is the SNR at UE $j$ (Since at most one BS transmits at any time slot, SINR is replaced by SNR). Clearly, the optimization problem of (\ref{Eq: pp BS indep optimization}) is convex and 
can be 
solved easily. Note that in this optimization we solve the one-time transmit power for all UEs. The same UE might be selected by the corresponding BS in multiple blocks, but the transmit power for that UE stays unchanged. In this regard, we ignore the block index of the transmit powers in (\ref{Eq: pp BS indep optimization}) and simply write $p_{j,i}(k,n)$ as $p_{j,i}(k)$. Then the objective function (\ref{Eq: pp BS indep optimization}) becomes  
\be 
\label{Eq: pp BS indep optimization, one time}
Z_{i}(k)    \left( p_{j,i}(k) \sum_{n\in[N]}     
\widetilde{T}^{\rm d}_{j,i}(k,n) -Tp_i^{\rm avg} \right) -H_{j,i}(k)\widehat{X}_{j,i}(k),
\ee
from which the transmit power $p_{j,i}(k)$ for each UE can be solved at the beginning of the epoch $k$. Similarly, to solve auxiliary variables, each BS needs to independently maximize $VU(\gamma_{j,i}(k))-H_{j,i}(k)\gamma_{j,i}(k)$  subject to $0\le \gamma_{j,i}(k)\le TW\log(1+g_{j,i}^{\rm max}p_i^{\rm max})$, which is also  convex.

In the $p$-persistent protocol, we let the BSs compete for the channel use in each block within each epoch.\footnote{{The reason that we let the channel contention happen in each block instead of each epoch is for the consideration of data transmission delay of the UEs. If one BS wins the channel contention and occupies it for the entire epoch, then all other BSs have to wait until the next epoch begins to contend again. This will result in a significant delay for other UEs since the length of an epoch could be much longer than a block.}} To avoid interference, there can be at most one  active link at any time. More specifically, at the beginning of each block, each BS attempts to transmit with probability $P_{\rm c}$. If more than one BS decide to transmit at the same time, i.e., collisions are detected, then all BSs will not transmit. The BSs then contend the channel again in the following time slot until one BS wins the channel, i.e., there is only one BS decides to transmit and all other BSs stay silent. The BS which wins the contention then randomly chooses one UE from the set of UEs associated with it to transmit to it until the end of the current block. All BSs will contend for the channel again at the beginning of the next block. At any time slot, successful transmission happens with probability $MP_{\rm c}(1-P_{\rm c})^{M-1}$ which is maximized when $P_{\rm c}=1/M$. 
Note that the above channel contention process can also be used as a simulated process which produces an estimation for the data transmission time for the UEs during the current epoch. 

\subsubsection{CSMA/CA Access Strategy} 
We consider a CSMA/CA MAC protocol with exponential backoff time (IEEE 802.11). Different from the $p$-persistent case, the CSMA/CA scheduling happens in each epoch instead of in each block. More specifically, each BS listens to the shared spectrum before transmitting. If the channel is sensed to be busy, the BS will wait. If the channel is idle, the BS starts to transmit to its selected UE with certain probability.  
If a collision occurs, each BS then chooses a random backoff time of 1 or 2 slots (assuming a contention window size of two) and attempts to transmit again after the chosen backoff time. If no collision occurs, the BS wining the channel in the last slot will randomly choose a backoff time of 1 or 2. If collision happens again, each BS randomly chooses a backoff time between 1, 2, 3 and 4. After $C$ collisions, each BS will choose a backoff time randomly distributed from 1 to $2^{C}$ and attempts to transmit again after the chosen backoff time. The maximum backoff time can not exceed the epoch length $T$. To improve the data transmission efficiency, a BS wining the channel contention may continue its data transmission for multiple consecutive slots  instead of only one. Similar to the case of the $p$-persistent MAC, at the beginning of each epoch, based on an estimation of the data transmission time for each UE, each BS independently solves the sub-problem (\ref{Eq: pp BS indep optimization}). Because there is only one active link at any time, independent optimizations performed by the individual BSs are equivalent to the joint optimization of the sub-problems (\ref{Eq: sub-opt 1}) and (\ref{Eq: sub-opt 2}) as in the case of $p$-persistent MAC. Note that the transmit power for each UE is determined by solving the second sub-problem at the beginning of each epoch and will stay unchanged during the entire epoch. We further assume that the UE selection of the BSs is fixed during each epoch but can change among different epochs. Particularly, at the beginning of each epoch, we let each BS randomly select one of its associated UEs to serve throughout the whole epoch, i.e., at any slots in which the BS wins the channel contention.

\subsubsection{The Ideal Case} 
\label{Section: ideal case}
To give a straightforward intuition on the optimality of the proposed scheduling algorithm, we consider a scenario in which we assume there is no interference among the BSs. In particular, at the beginning  of each epoch, each BS $i$ randomly selects a  UE $j(i)\in\mathcal{K}_i$ to serve throughout the whole epoch. The $M$ BSs then transmit to its selected UEs simultaneously and there is no interference among them. Note that this `ideal case' is just a way to produce an upper bound on the performance and is not an achievable scheme. 
Since in this case the data transmission time for each UE can be easily determined at the beginning of each epoch, the transmit powers (and the auxiliary variables) of the BSs can be determined by solving the sub-problems  (\ref{Eq: sub-opt 1}) and (\ref{Eq: sub-opt 2}) similarly to that of the  $p$-persistent and CSMA/CA protocols. 

\subsubsection{$p$-persistent and CSMA/CA Protocol with Random Powers}
\label{Section: PP and CSMA with random power allocation}
We provide two more baselines which are the $p$-persistent and CSMA/CA protocols with random transmit powers. In particular, the two protocols are preformed repeatedly at each epoch and the BS transmit powers are chosen randomly. Due to the choice of the peak and average power constraints, the average power constraints of the BSs can be satisfied.


\subsection{Simulation Result}
In this section we present the numerical results on the performance of the proposed game-based scheduling algorithm. We compare the performance of the proposed approach to the baseline schemes, i.e., the $p$-persistent and CSMA/CA MAC protocols with optimized and random transmit powers. The simulation setup is described as follows.

\begin{figure}[ht]
\centering
\includegraphics[width=0.5\textwidth]{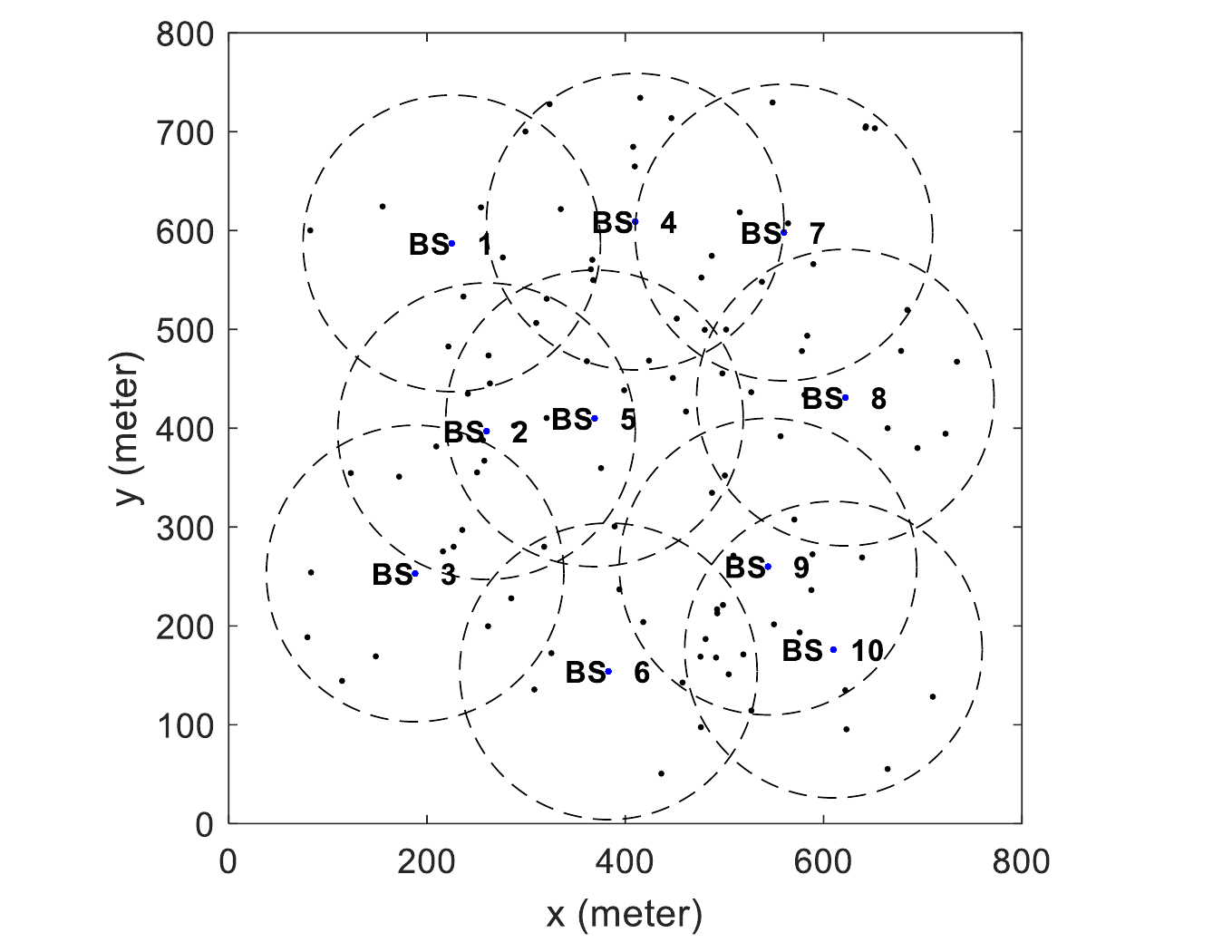}
\caption{\small Distribution of the BSs and the UEs over a planar grid of dimension $800\times 800$ meter squared. Each BS is associated with ten UEs and these UEs are uniformly distributed in the disk coverage area with a radius of 150 meters. This BS-UE association is fixed over all epochs.}
\label{Fig: BS UE positions}
\end{figure}

Consider a wireless network (See Fig.~\ref{Fig: BS UE positions}) with $M=10$ BSs, each from a different operator, and a total of $K=100$ UEs located on a planar grid with dimension $800\times 800$ meter squared. Each BS is responsible for serving a set of $10$ UEs within its coverage area which is a disk of radius 150 meters centered at that BS. Due to the proximity locations, the coverage areas of  different BSs may overlap. The system operates on a  $W=400$ MHz frequency band with a center frequency $W_{\rm c}=37$ GHz. Each BS $i$ has an average power constraint of $p_i^{\rm avg}= 38.13$ dBm ($ 6.5$ Watt) and a peak power of $p_i^{\rm max}=39$ dBm (7.9 Watt). For the wireless propagation channels, the path loss factor is set to be $\eta=4$. The parameters of the Nakagami-m distribution are $\mu=1,\Omega=0.001$. Each time slot represents 1 millisecond. Each block contains $T^{\rm b}=50$ slots and each epoch contains $N=8$ blocks thus having $T=NT^{\rm b}=400$ slots. 
For the $p$-persistent baseline scheme, the optimal contention probability is set to be $P_{\rm c}=0.1$. For the CSMA/CA scheme, the minimum contention window is set to be ${\sf CW}_{\rm min}=20$ slots. For practical reasons, we also impose a maximum contention window constraint of ${\sf CW}_{\rm max}=200$ slots. Each data transmission duration contains two time slots. The random noise power at the UEs is calculated according to
\be 
\sigma^2 \,(\textrm{dBm}) = 10\lg\left(k_{\rm B}T_0\times 10^3\right) + \textrm{NR}\,\textrm{(dB)} + 10\lg W, 
\ee
where $k_{\rm B}=1.38\times 10^{-23}$ Joules/Kelvin is the Boltzmann's constant, NR is the UE noise figure and $T_0$ is the temperature of UE receive antenna system. Taking the typical value of $\textrm{NR}=1.5\, \textrm{dB}$ and $T_0 = 290\, \textrm{Kelvin}$, the total noise power over the 400  MHz bandwidth is equal to $\sigma^2 =-86.46 \,\textrm{dBm}$.
In the simulation, we also assume that the BSs' and UEs' beams  are perfectly aligned, i.e., if a UE is served by a BS, then the UE will lie in the center of the BS antenna main-lobe and the BS will lie in the center of the UE antenna main-lobe. 
With the above parameter specification, we next evaluate the performance of the proposed game-based scheduling algorithm and verify the effect of BS/UE beam width,  MSR, the number of UEs, and feedback overhead on the network utility. In all simulations, we fix the Lyapunov constant to be $V= 1000$.

\subsubsection{Effect of BS/UE Beam Width} The BS MSR is fixed as $D^{\rm BS}=20 \textrm{ dB}$. We let the BS beam width take values $\Delta\theta^{\rm BS}=\frac{\pi}{9},\frac{\pi}{36}$ and $\frac{\pi}{72}$ respectively in order to verify the effect of the beam width. Since changing the UE antenna beam width and MSR has a similar effect as varying that of the BSs, we fix the UE antenna beam width and the MSR to be $\Delta\theta ^{\rm UE}=\frac{\pi}{18}$ and  $D^{\rm UE}=10 \textrm{ dB}$.
\begin{figure*}[t]
\centering
\subfigure[BS beam width $\Delta\theta^{\rm BS}={\pi}/{9}$,  $D^{\rm BS}=20 \textrm{ dB}$.]{\includegraphics[width=0.32\textwidth]{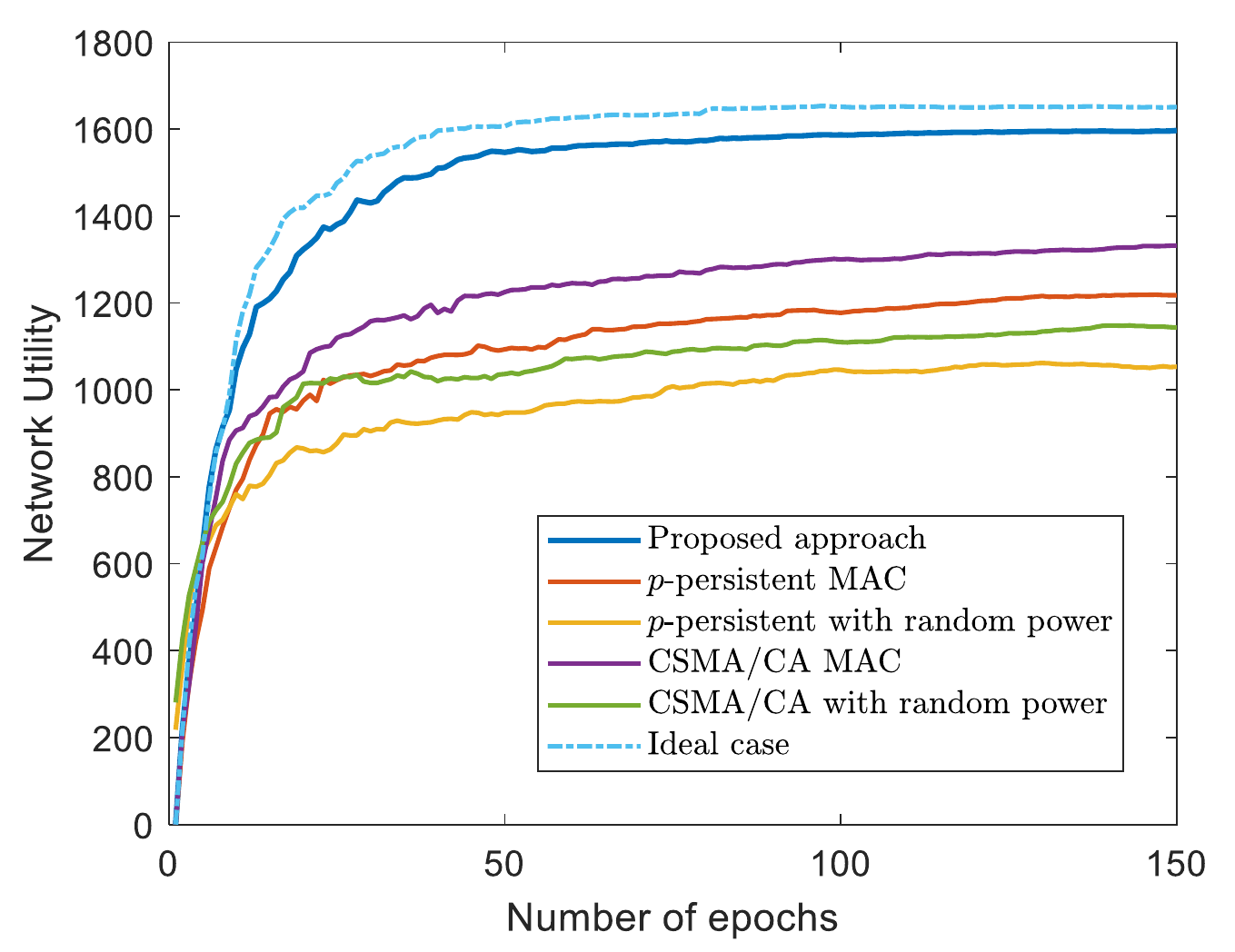}}
\subfigure[BS beam width $\Delta\theta^{\rm BS}=\pi/36$, $D^{\rm BS}=20 \textrm{ dB}$.]{\includegraphics[width=0.32\textwidth]{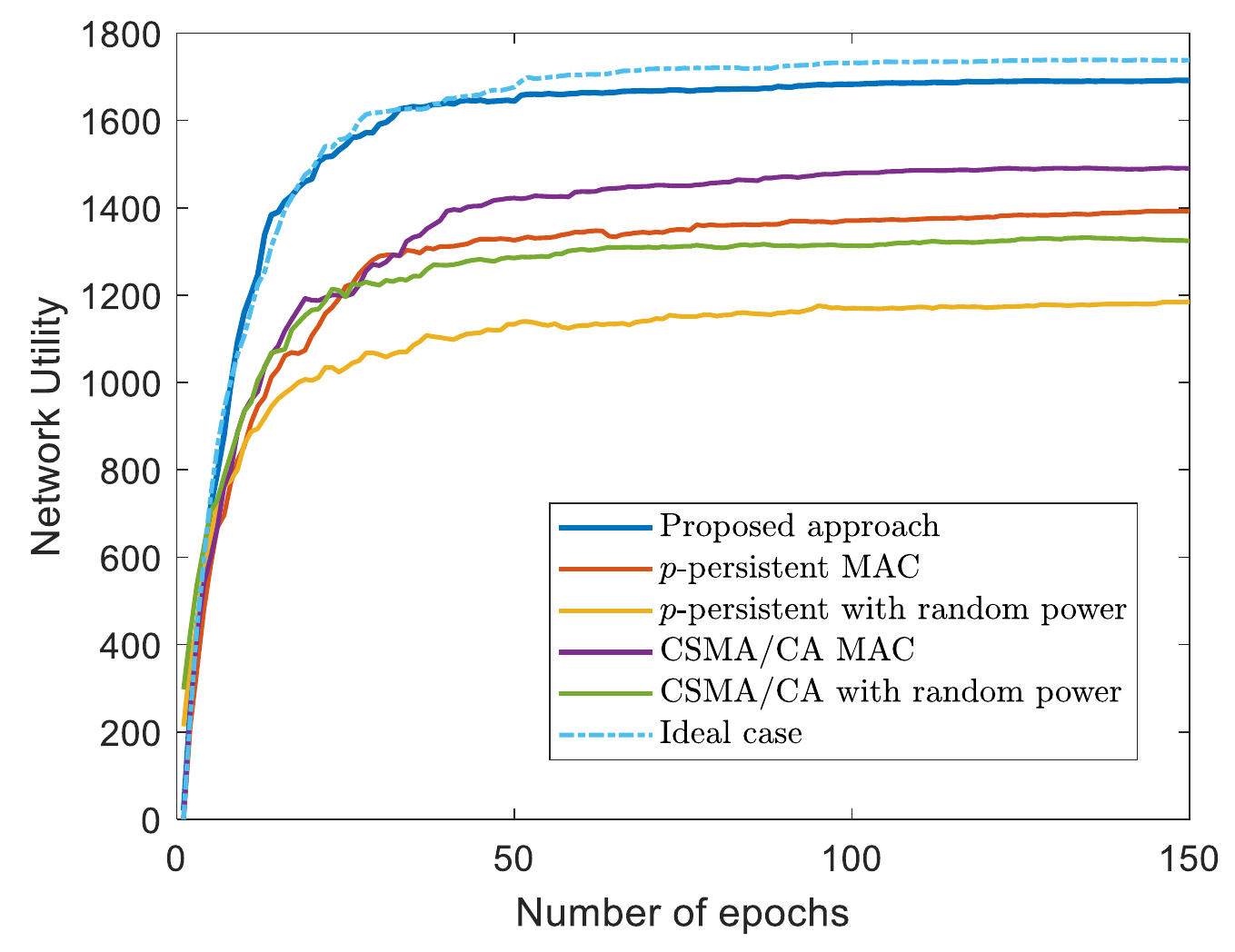} }
\subfigure[BS beam width $\Delta \theta^{\rm BS}=\pi/72$,  $D^{\rm BS}=20 \textrm{ dB}$.]{\includegraphics[width=0.32\textwidth]{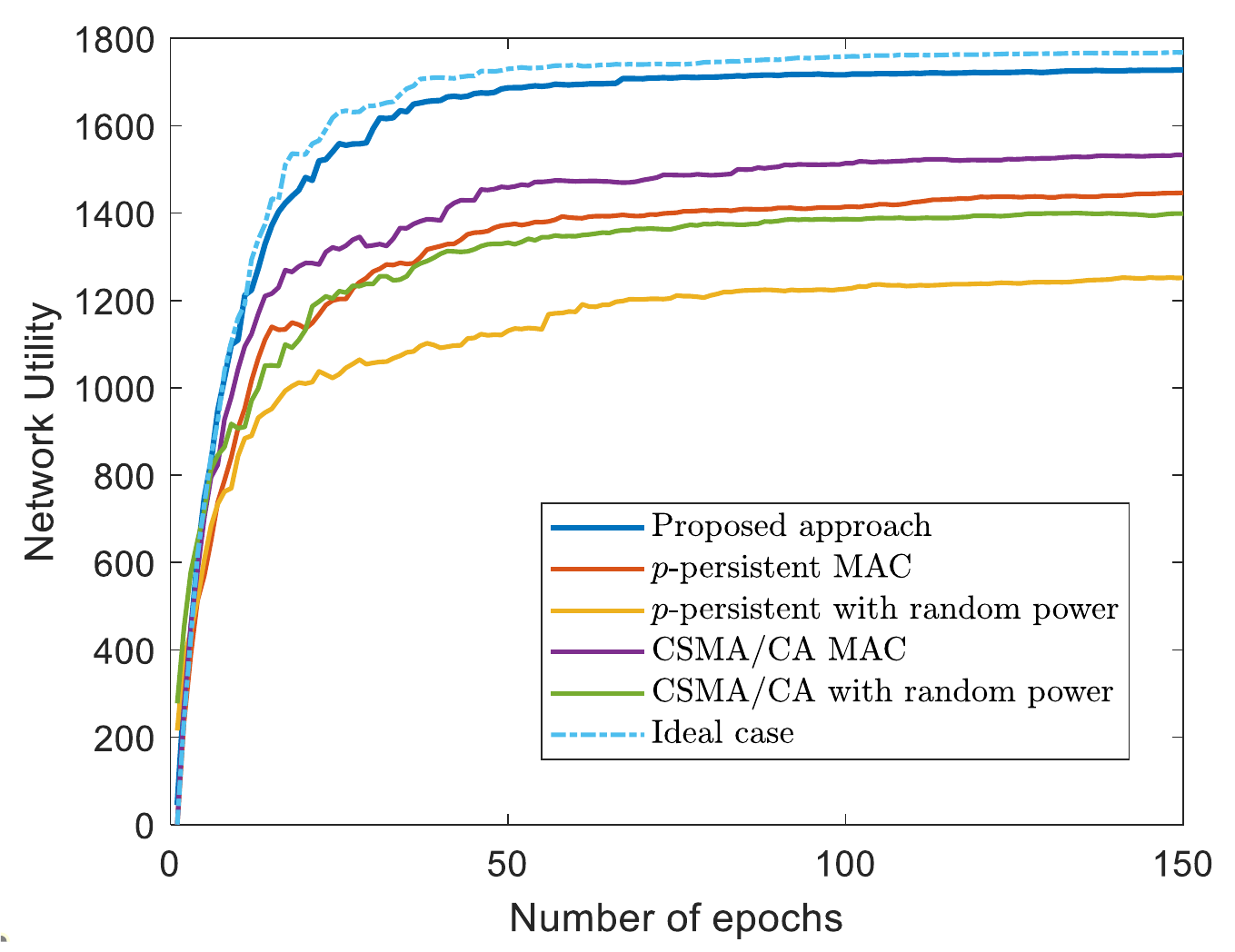} }
\vspace{-0.2cm}
\caption{\small The effect of BS beam width $\Delta \theta^{\rm BS}$ on the network utility. The BS antenna MSR is fixed to be $D^{\rm BS}=20 \textrm{ dB}$.}
\label{Fig: effect of beam width}
\end{figure*}
\begin{figure*}[t]
\centering
\subfigure[Proposed approach under different values of BS beam width $\Delta\theta^{\rm BS}$, $D^{\rm BS}=20 \textrm{ dB}$.]{\includegraphics[width=0.32\textwidth]{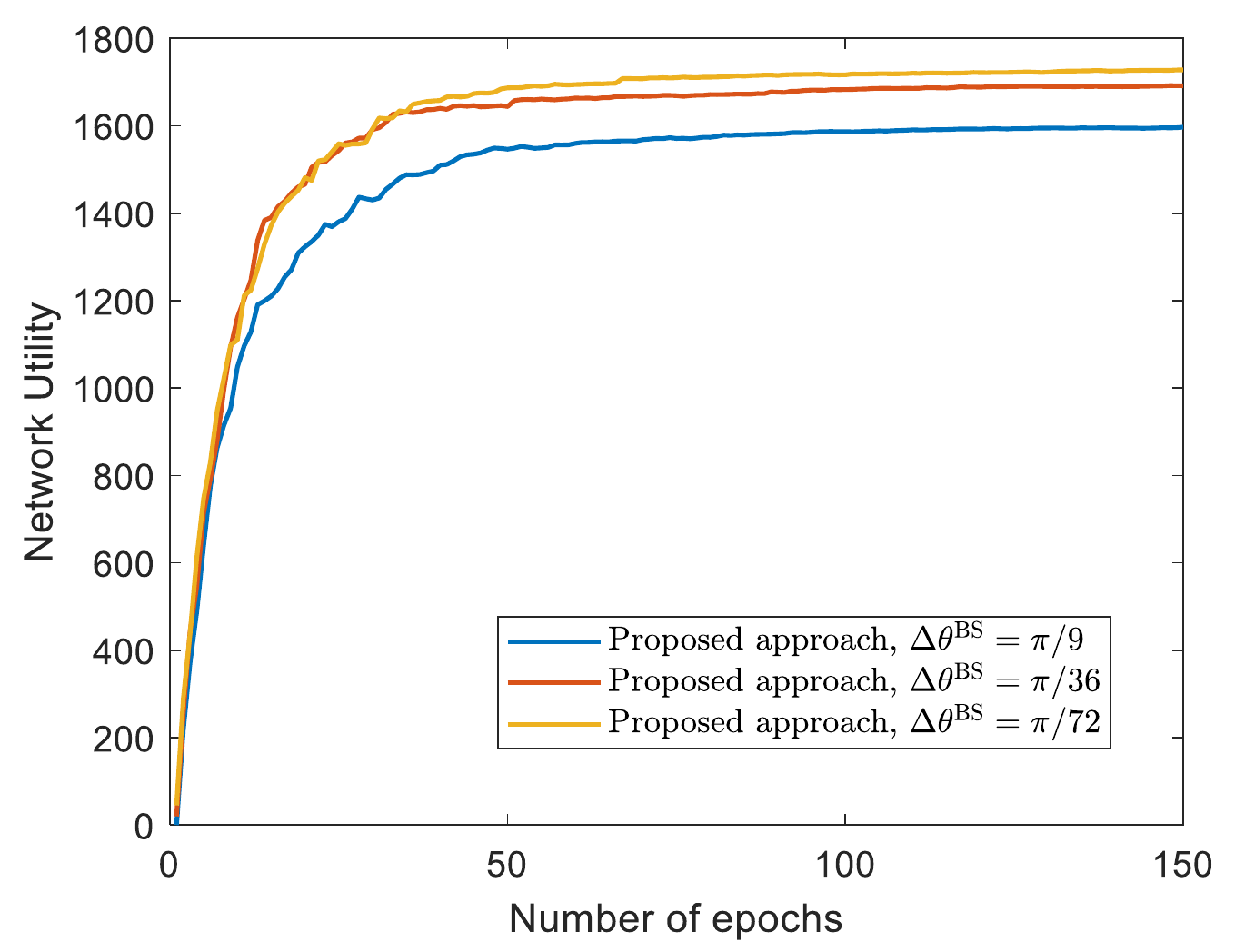}}
\subfigure[$p$-persistent strategy under different values of BS beam width $\Delta\theta^{\rm BS}$, $D^{\rm BS}=20 \textrm{ dB}$.]{\includegraphics[width=0.32\textwidth]{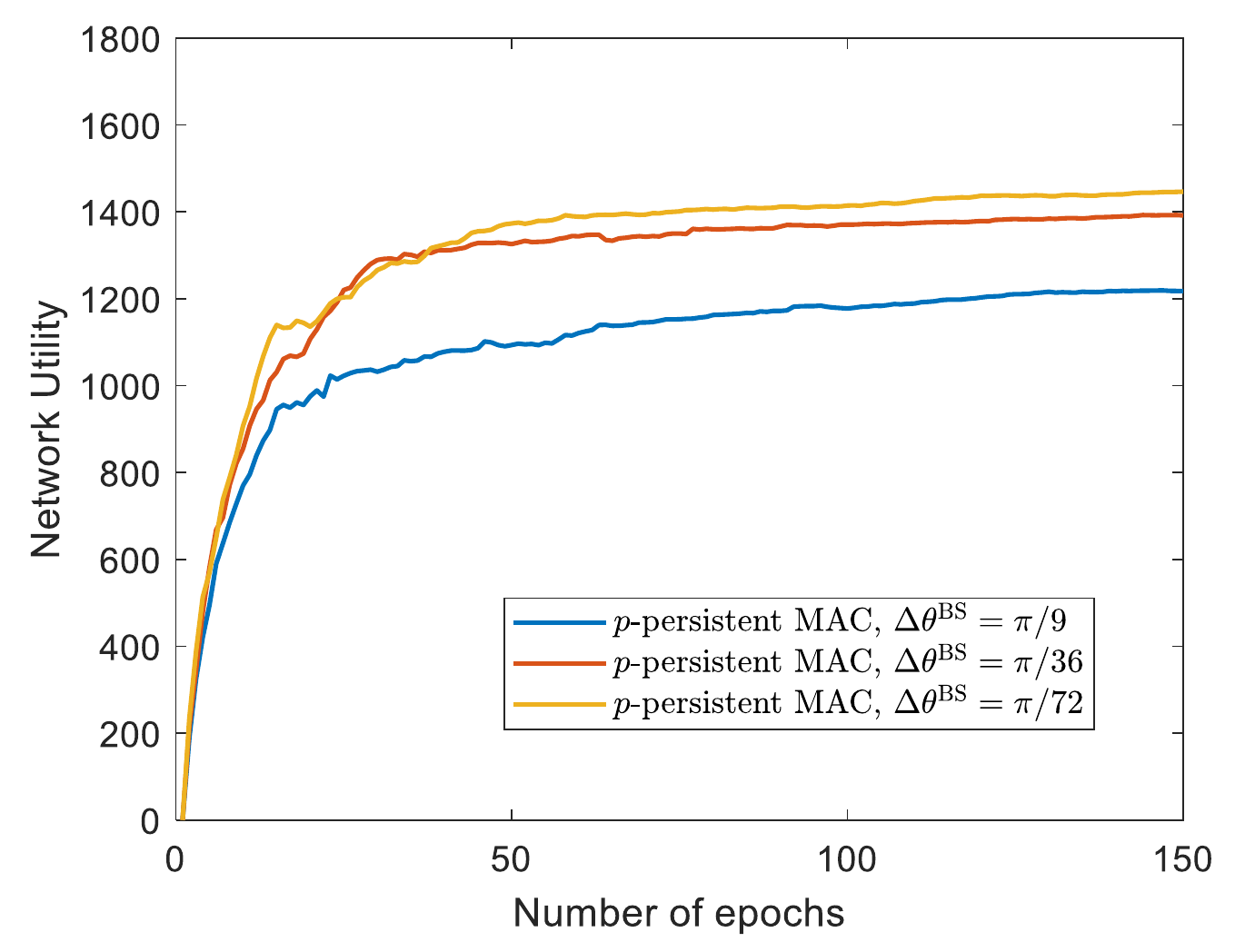}}
\subfigure[CSMA/CA strategy under different values of BS beam width $\Delta\theta^{\rm BS}$, $D^{\rm BS}=20 \textrm{ dB}$.]{\includegraphics[width=0.32\textwidth]{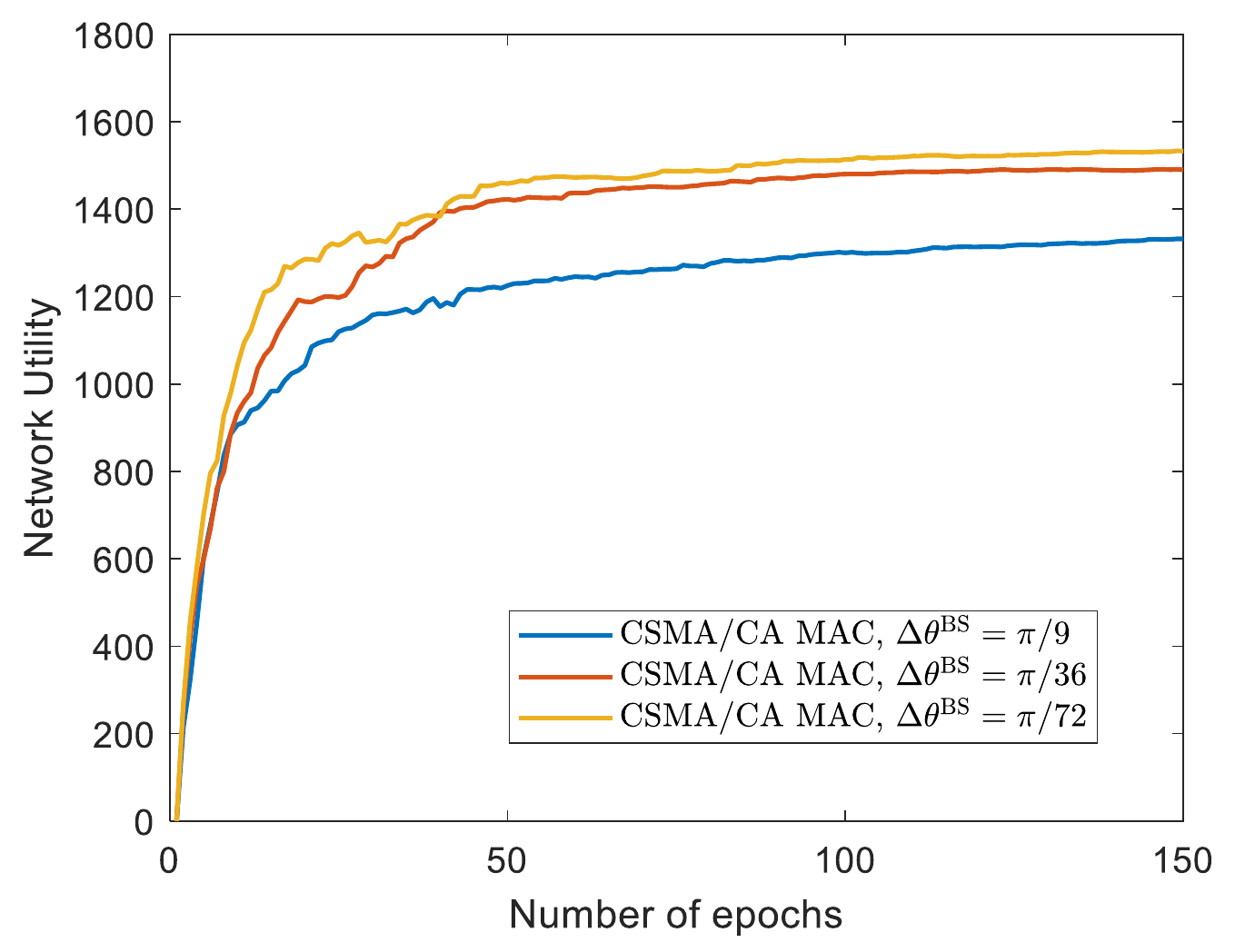} }
\vspace{-0.2cm}
\caption{\small The effect of BS beam width on the network utility for different access schemes. The BS antenna MSR is fixed to be $D^{\rm BS}=20 \textrm{ dB}$.}
\label{Fig: comparison verus beam width}
\end{figure*}
The network utility versus the number of epochs curve is shown in Fig.~\ref{Fig: effect of beam width}. We have the following observations. First, for all the three cases, the proposed approach outperforms the baseline schemes. More specifically, the proposed approach converges faster than the baselines and achieves higher asymptotic utility. It can also be seen that the CSMA/CA baseline achieves better utility than the $p$-persistent baseline for both optimized and random transmit powers. This is because in the CSMA/CA protocol, the data transmission time has been improved compared to the $p$-persistent protocol. Second, it can be seen that when the beam becomes narrower, the achieved network utility of all schemes increase (See Fig. \ref{Fig: comparison verus beam width}). This is because narrower BS beams increase the antenna gain towards the target UE and avoid covering other UEs and causing interference. Note that when the BS antenna beam width is very small and the MSR $D^{\rm BS}$ is very large, the proposed approach will have a similar performance to the ideal case since very sharp beams will eliminate the interference from undesired BSs for the UEs and mimic the performance of the ideal case in which it is assumed that BSs do not interfere with each other.

\subsubsection{Effect of BS/UE MSR} 
\begin{figure*}[t]
\centering
\subfigure[BS MSR $D^{\rm BS}=10\textrm{ dB}, \Delta\theta^{\rm BS}=\pi/18$.]{\includegraphics[width=0.321\textwidth]{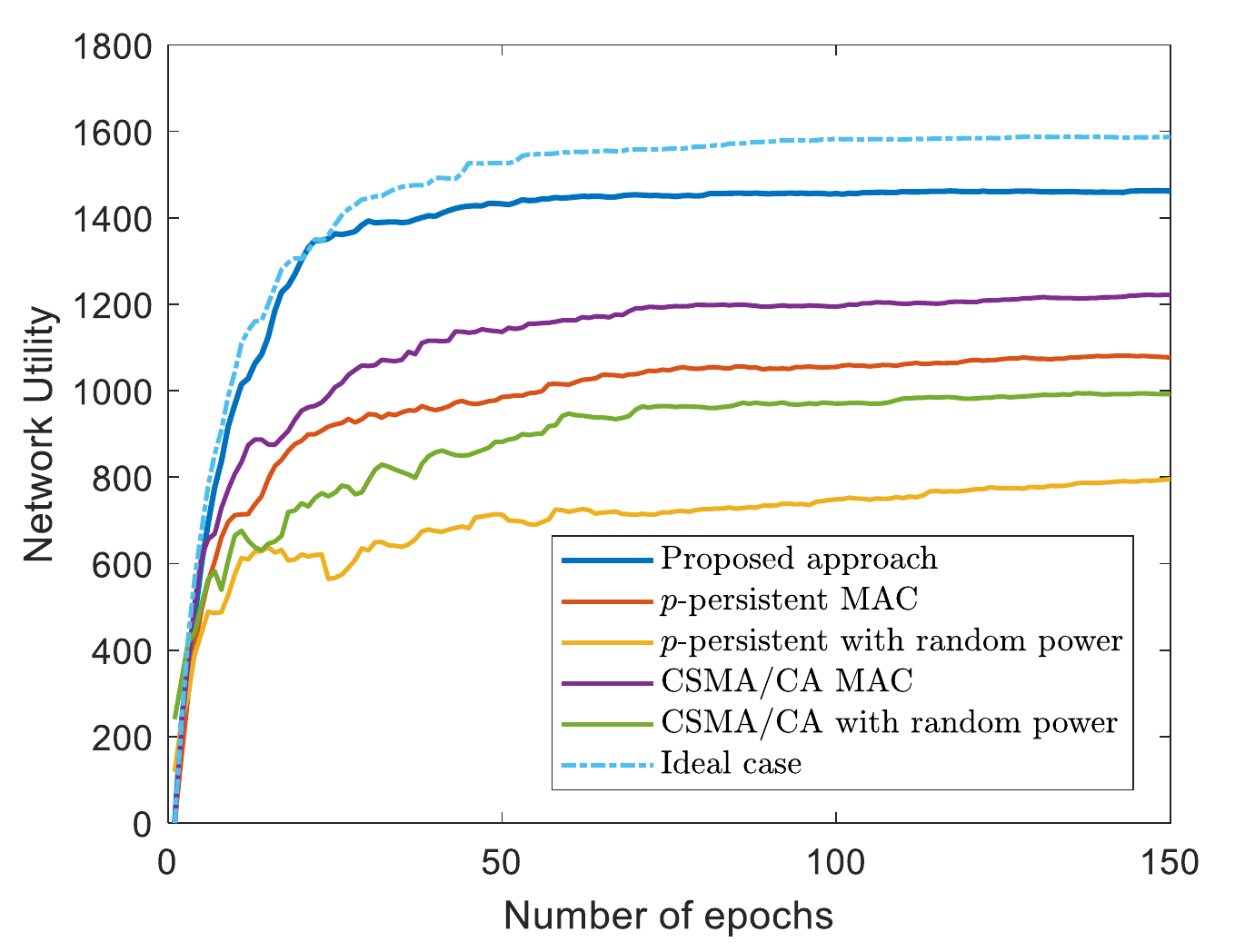}}
\subfigure[BS MSR $D^{\rm BS}=20\textrm{ dB}, \Delta \theta^{\rm BS}=\pi/18$.]{\includegraphics[width=0.321\textwidth]{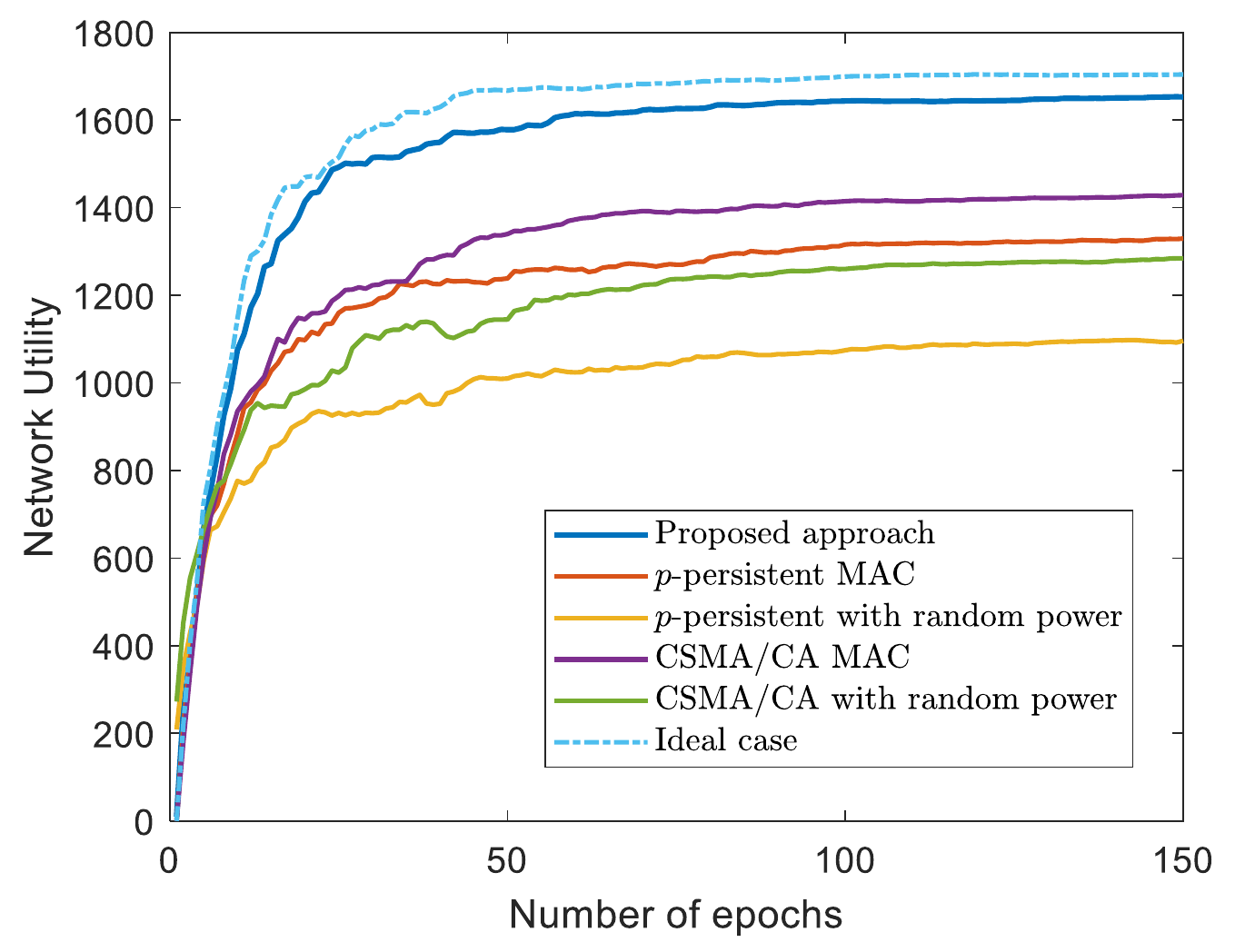}}
\subfigure[BS MSR $D^{\rm BS}=30\textrm{ dB}, \Delta \theta^{\rm BS}=\pi/18$.]{\includegraphics[width=0.321\textwidth]{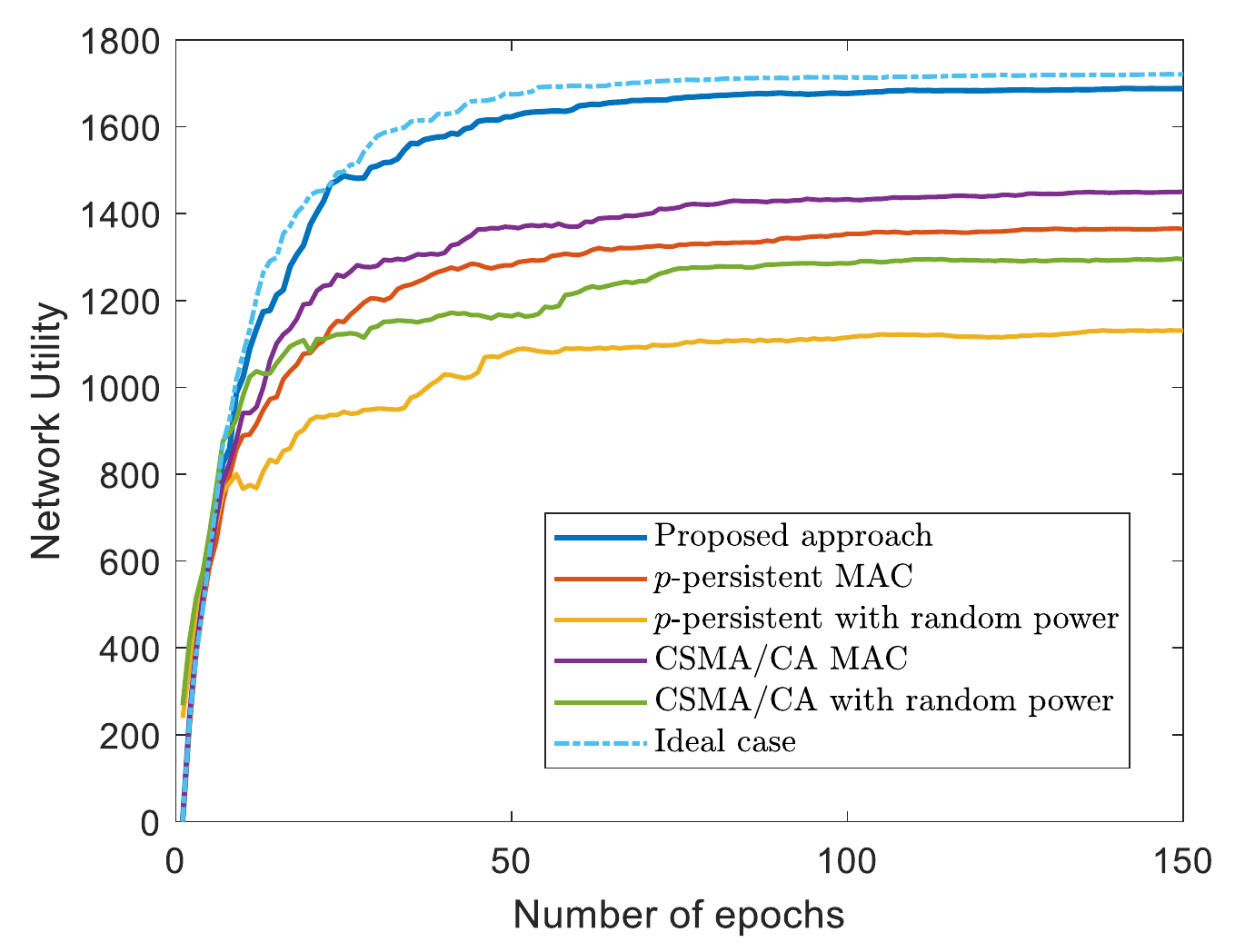}}
\vspace{-0.2cm}
\caption{\small The effect of BS MSR on the network utility. The BS beam width is fixed to be $\Delta\theta^{\rm BS}=\pi/18$.}
\label{Fig: effect of directivity}
\end{figure*}

\begin{figure*}[t]
\centering
\subfigure[Proposed approach under different values of BS MSR $D^{\rm BS},\Delta\theta^{\rm BS}=\pi/18$.]{\includegraphics[width=0.32\textwidth]{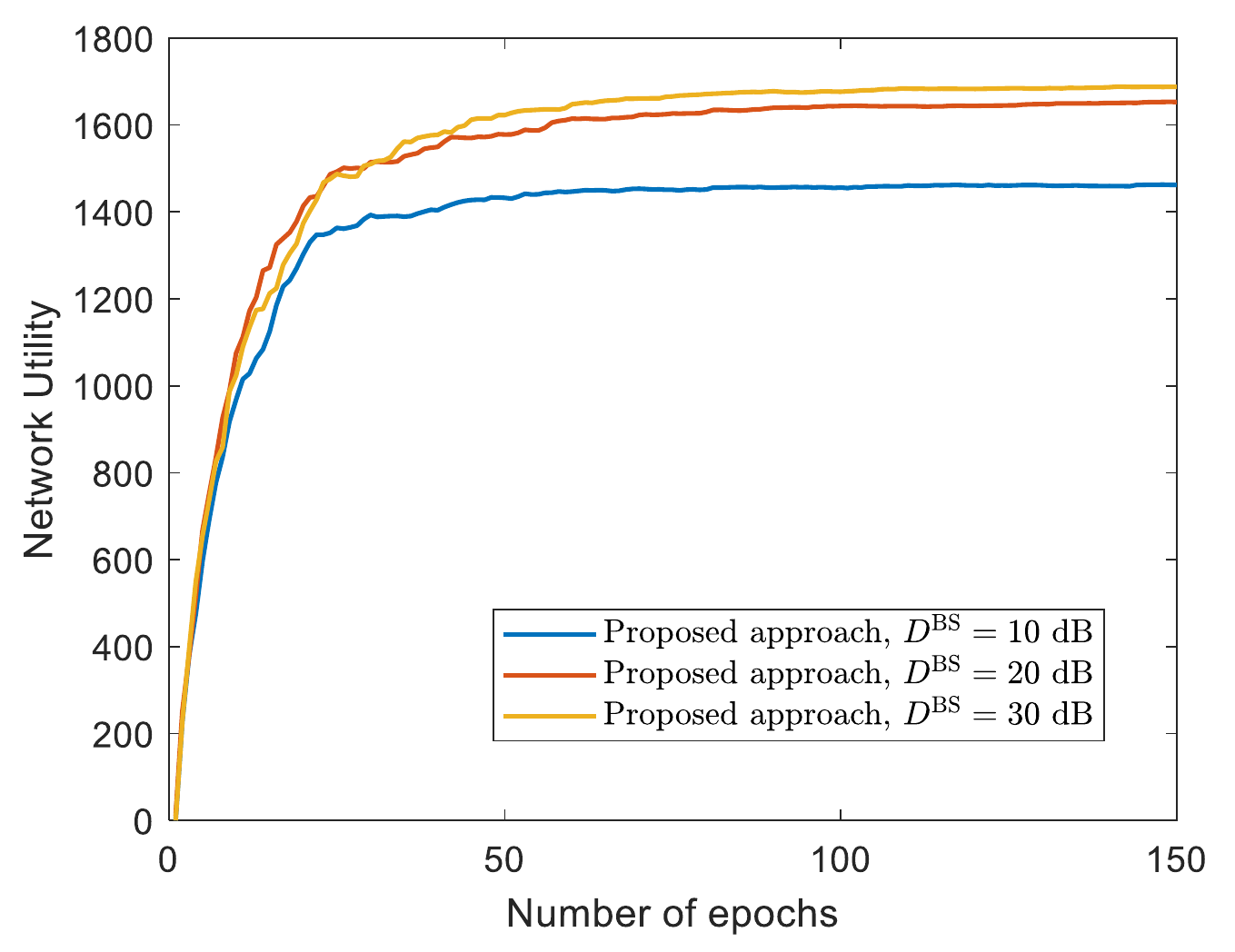}}
\subfigure[$p$-persistent strategy under different values of BS MSR $D^{\rm BS},\Delta\theta^{\rm BS}=\pi/18$.]{\includegraphics[width=0.32\textwidth]{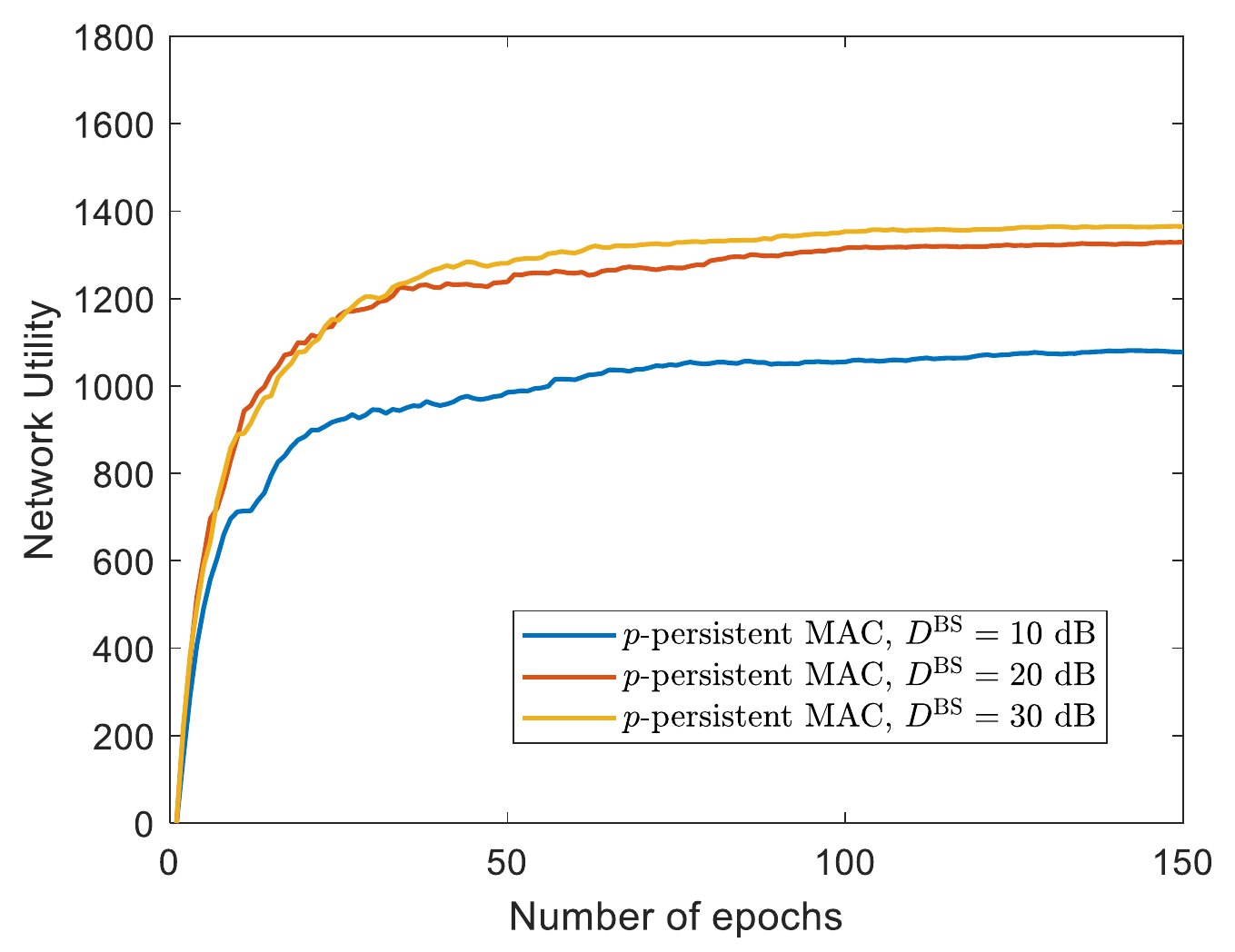}}
\subfigure[CSMA/CA strategy under different values of BS MSR $D^{\rm BS},\Delta\theta^{\rm BS}=\pi/18$.]{\includegraphics[width=0.32\textwidth]{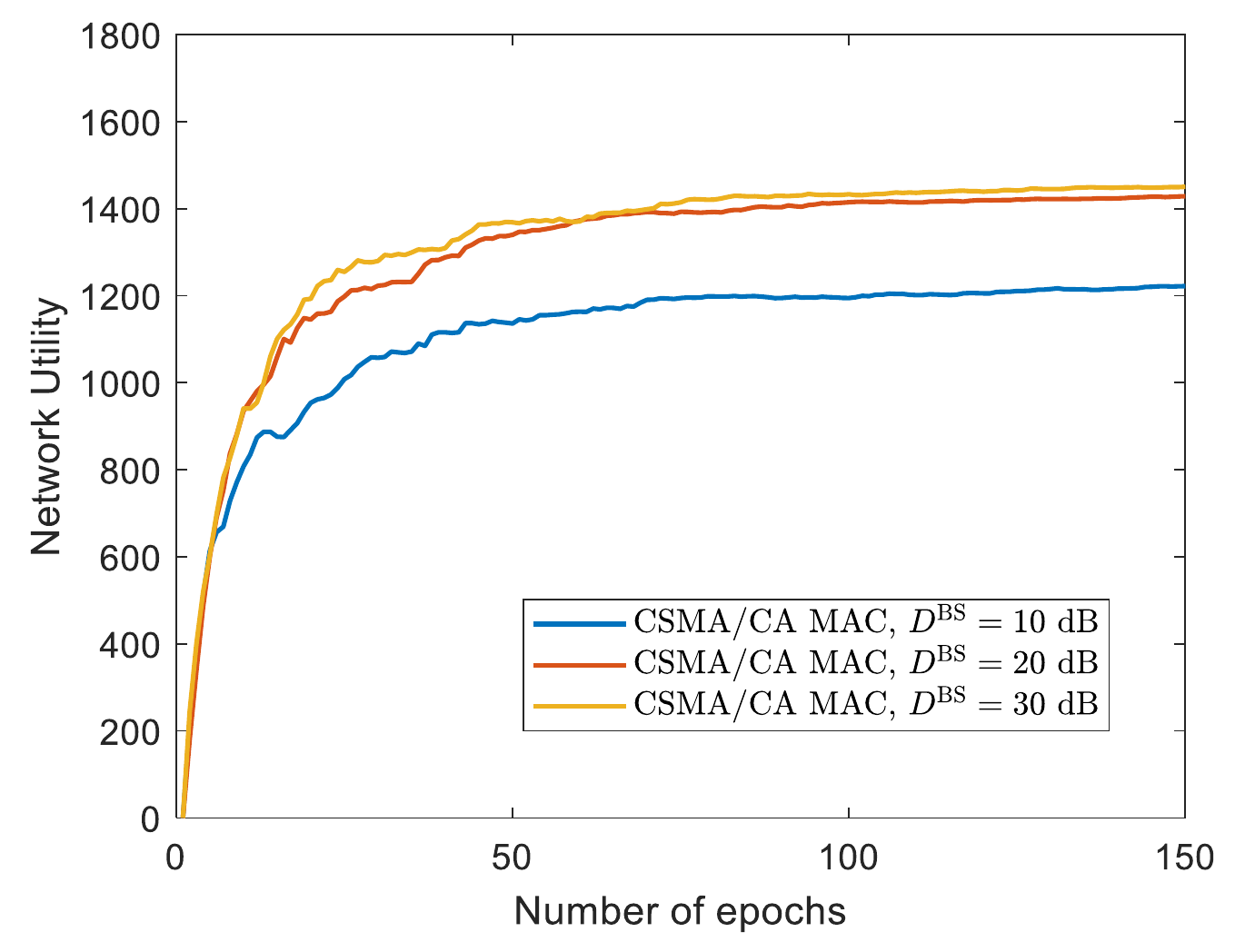} }
\vspace{-0.2cm}
\caption{\small The effect of the BS MSR $D^{\rm BS}$ on the network utility for different access schemes. The BS antenna beam width is fixed to be $\Delta\theta^{\rm BS}=\pi/18$.}
\label{Fig: comparison verus directivity}
\end{figure*}

The UE antenna beam width and MSR are fixed as $ \Delta\theta^{\rm UE}=\frac{\pi}{18}, D^{\rm UE}=10 \textrm{ dB}$. 
The BS antenna beam width is fixed to be $\Delta\theta^{\rm BS}=\frac{\pi}{18}$. We let the BS MSR be $D^{\rm BS}=10,20$ and 30 dB respectively in order to verify the effect of MSR. The simulated curves are shown in Fig.~\ref{Fig: effect of directivity}. We have the following observations. First, for all the three cases, the proposed scheme outperforms the baselines in terms of both convergence speed and asymptotic utility. Second, it can be seen that when the MSR increases, the achieved network utilities of all the schemes increase (See Fig.~\ref{Fig: comparison verus directivity}). This is because a higher $D^{\rm BS}$  increases the antenna gain towards the target UE and reduces the side-lobe gain which causes interference to other UEs.

\subsubsection{Effect of the Number of UEs}
\begin{figure}[ht]
\centering
\subfigure[Network utility of the proposed approach for different number of UEs.]{\includegraphics[width=0.4\textwidth]{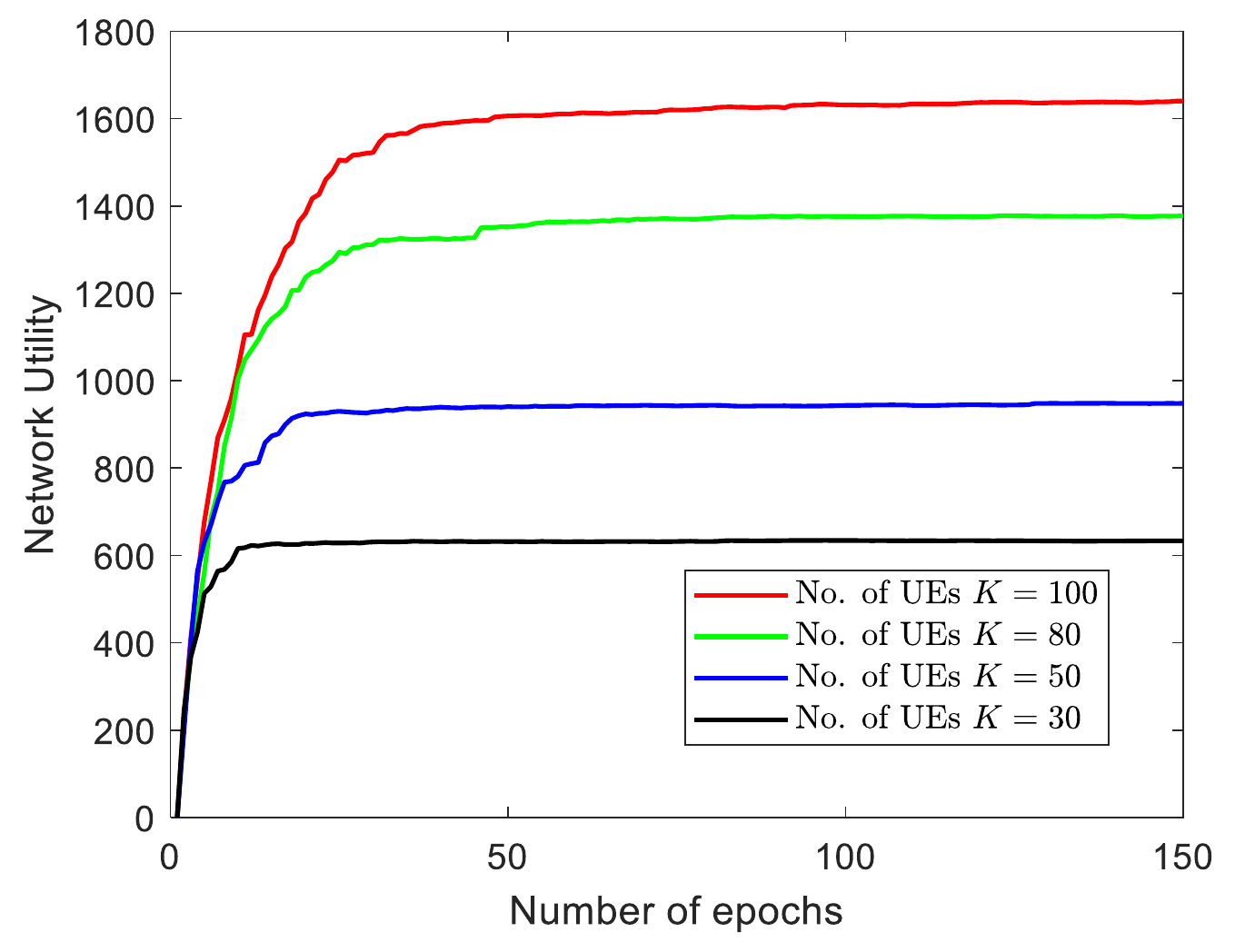}}
\subfigure[Average per UE utility of the proposed approach for different number of UEs.]{\includegraphics[width=0.4\textwidth]{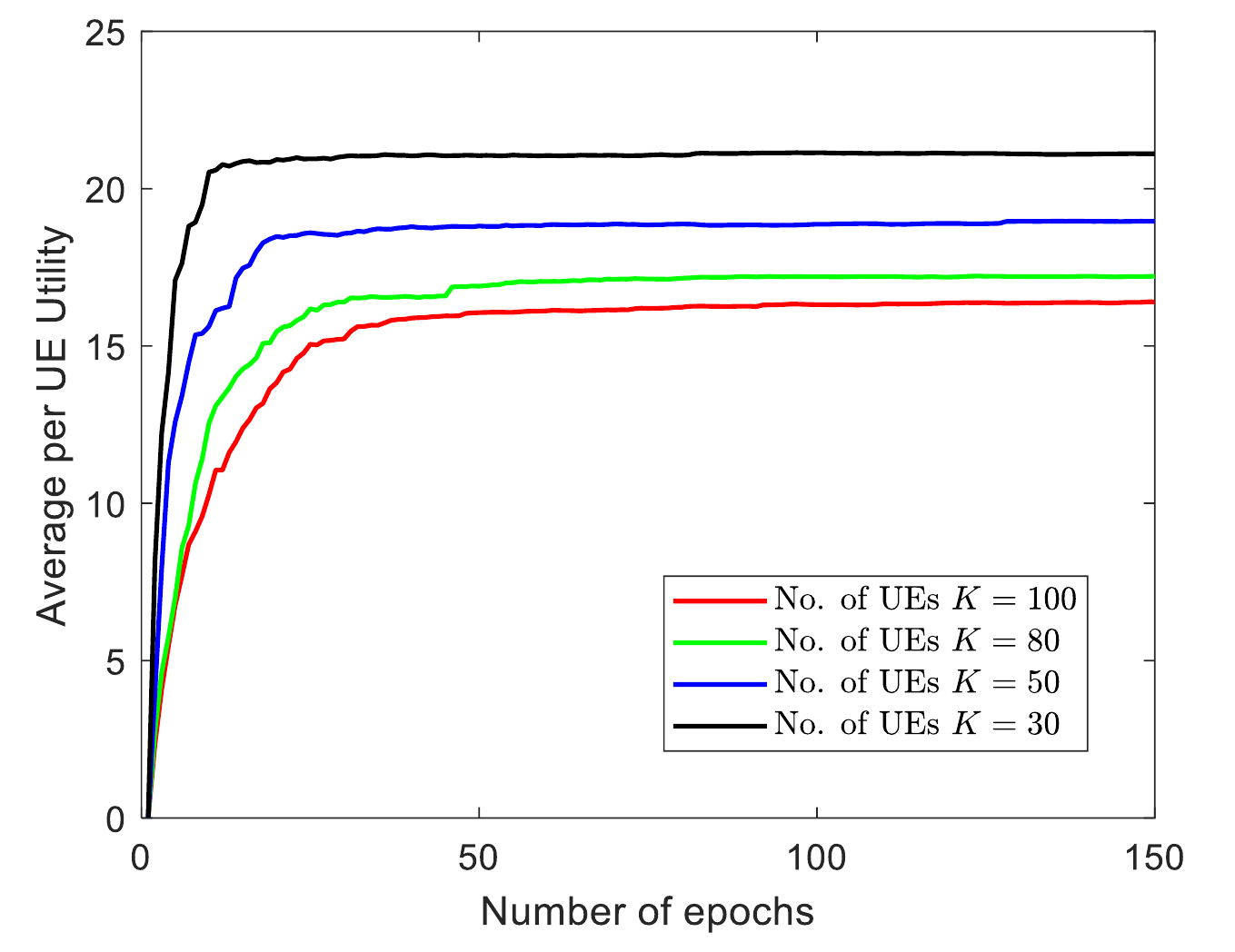}}
\subfigure[Network utility of the $p$-persistent MAC for different number of UEs.]{\includegraphics[width=0.4\textwidth]{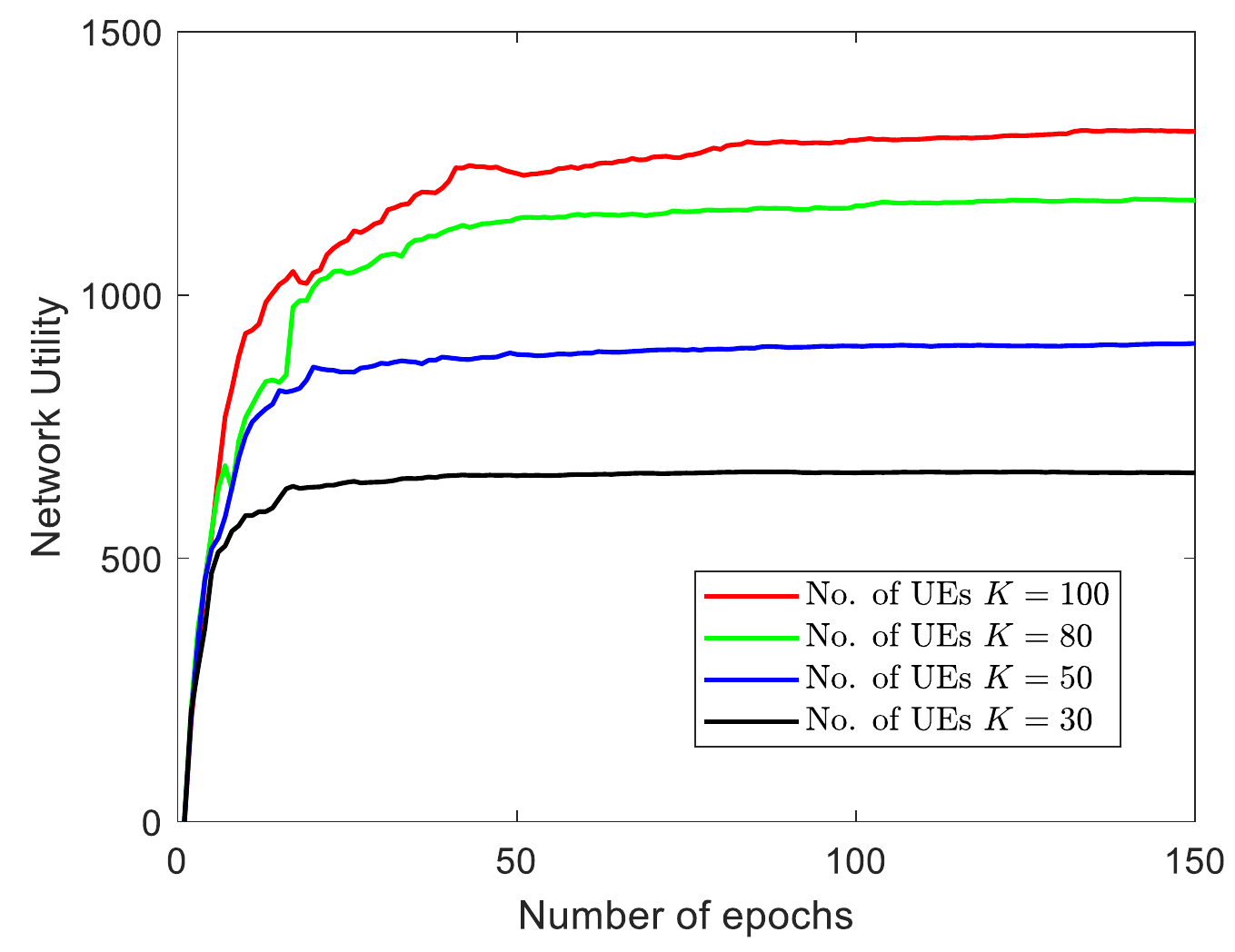}}
\subfigure[Network utility of the CSMA/CA MAC for different number of UEs.]{\includegraphics[width=0.4\textwidth]{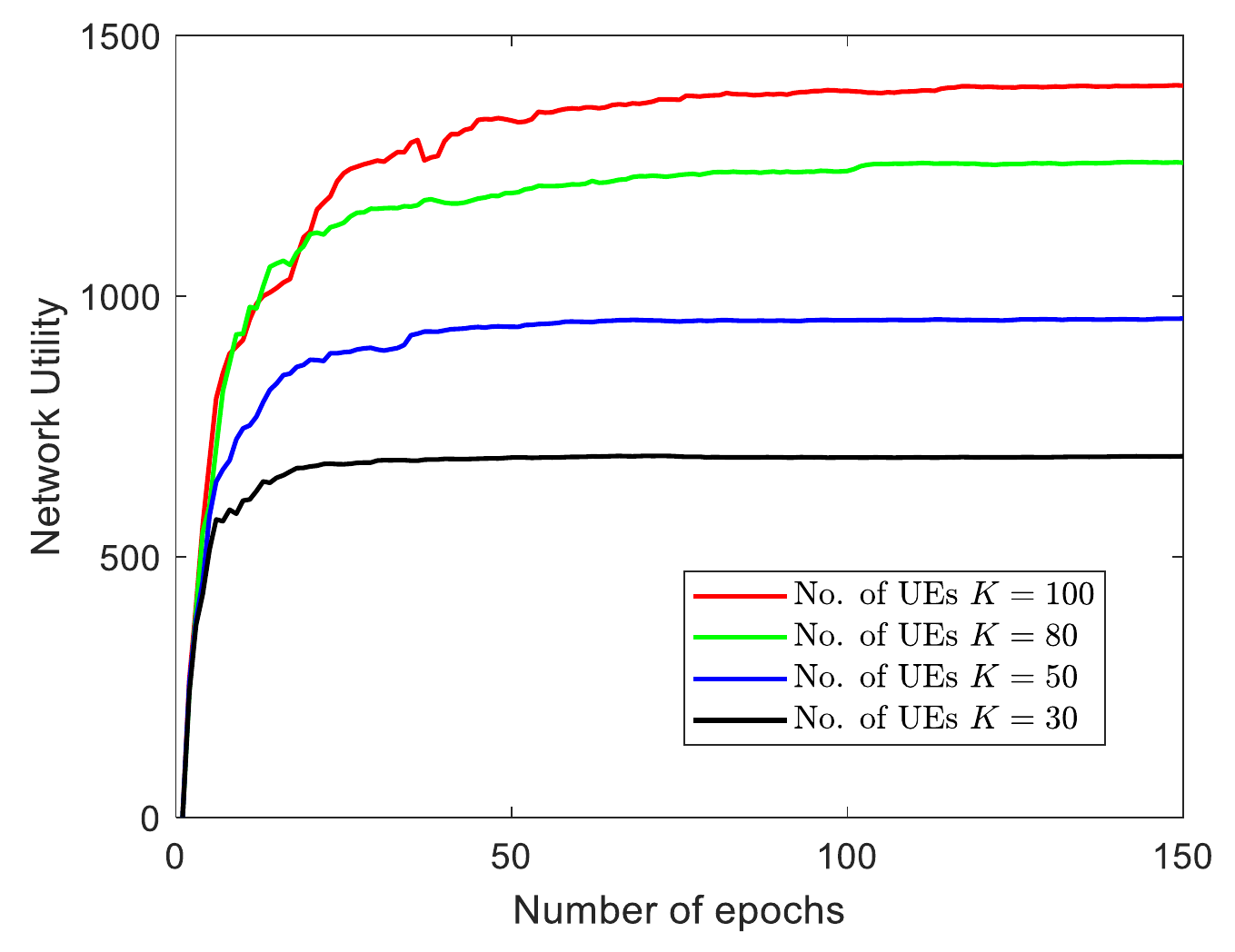}}
\vspace{-0.2cm}
\caption{\small The effect of the number of UEs on the network utility for the proposed approach and the baselines.}
\label{Fig: effect of no. of UEs}
\end{figure}

We verify the effect of the number of UEs on the proposed game-based scheduling algorithm in this section. The BS and UE antenna parameters are chosen as $\Delta \theta^{\rm BS }= \frac{\pi}{18},D^{\rm BS}=20 \textrm{ dB}$ and $\Delta \theta^{\rm UE}= \frac{\pi}{18},D^{\rm UE}=10 \textrm{ dB}$. The positions and coverage areas of the BSs are fixed as in Fig.~\ref{Fig: BS UE positions}. We then randomly generate $10,8,5$ and $3$ UEs for each BS. Therefore, the total number of UEs is equal to 100, 80, 50 and 30 respectively. The achieved network utility and average per UE utility, i.e., the network utility divided by the total number of UEs, are shown in Fig.~\ref{Fig: effect of no. of UEs}. From Fig.~\ref{Fig: effect of no. of UEs}(a), it can be seen that the achieved network utility increases as the number of UEs increases from 30 to 100. Similar trends can be observed from both baseline schemes as shown in Fig.~\ref{Fig: effect of no. of UEs}(c),(d). From Fig.~\ref{Fig: effect of no. of UEs}(b), it can be seen that the average per UE utility decreases as the number of UEs increases, and the utility curve converges faster (i.e., reach its asymptotic value in fewer epochs) when there are less UEs in the system. This result is expected because when the number of UEs increases, there are likely to be more UEs located close to the overlapping areas of the BS coverage. These UEs receives stronger interference from neighboring BSs and therefore achieves a smaller average utility.

\subsubsection{Optimality of the Proposed Scheduling Algorithm} 
\label{subsubsection: GT opt gap numerical evaluation}
\begin{figure}[ht]
\centering
\includegraphics[width=0.48\textwidth]{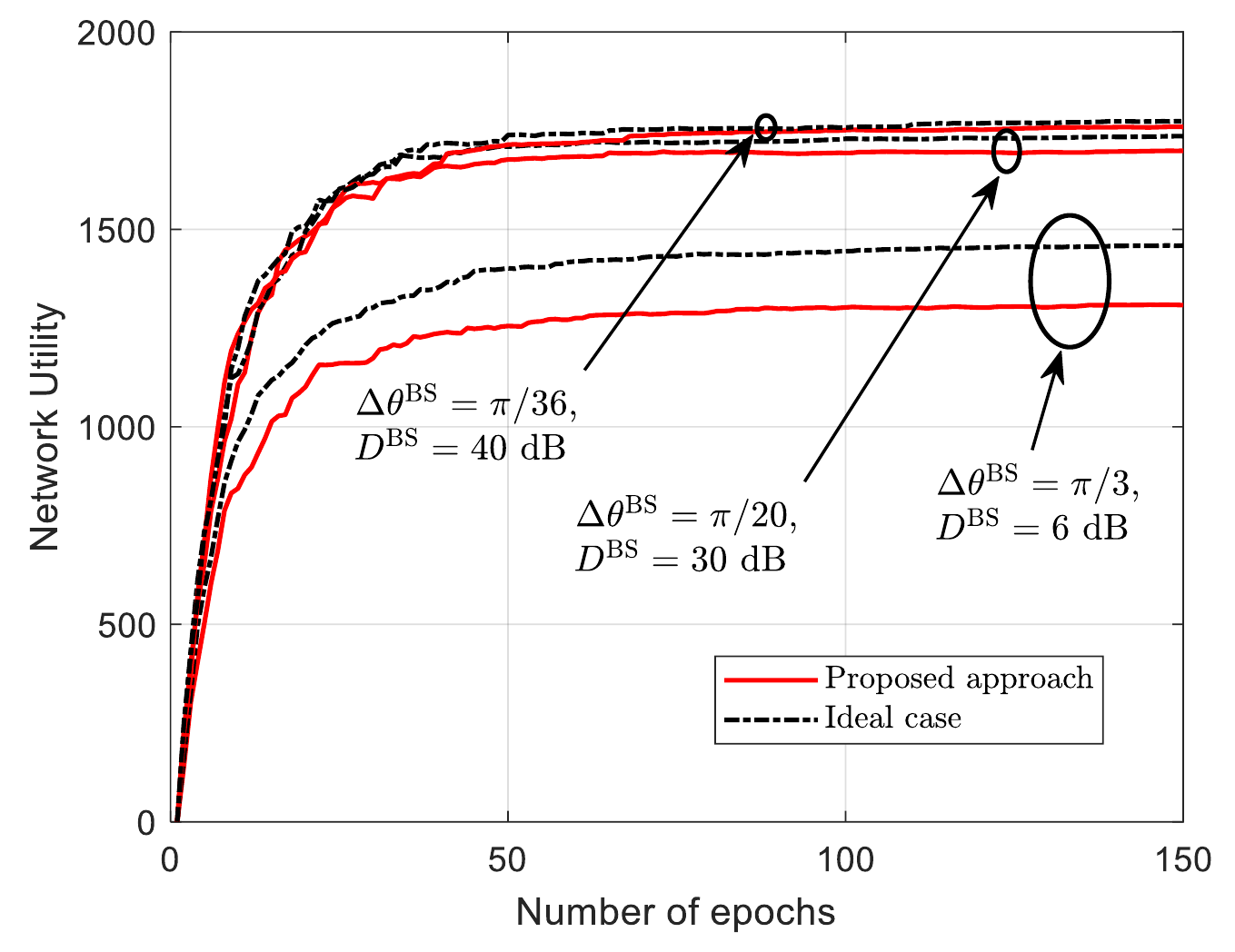}
\vspace{-0.2cm}
\caption{\small Optimality gap between the proposed approach and the ideal case for different antenna parameters.}
\label{Fig: GT opt gap}
\end{figure}
As we have observed in the previous simulation results, when the BS antenna beam becomes sharper, i.e., a narrower beam width and a larger MSR, the proposed game-based approach gets closer to the ideal case in terms of the achieved network utility. The reason is that, in the proposed algorithm, BSs update their transmit powers  based on the measured interference (plus noise) from all other BSs. When the BS  beam width $\Delta \theta^{\rm BS} $ is large, or the BS MSR $D^{\rm BS}$ is small, each UE is more likely to be covered by the main-lobe of many other interfering BSs, which will impose a strong interference to the UE and lead to performance degradation in throughput and therefore in network utility. In contrast, when the beams are sharp (small beam width and high MSR), the interference caused by neighboring BSs remains negligible as the side-lobe gain is very small, which is similar to the ideal case. Fig.~\ref{Fig: GT opt gap} shows the utility gap between the proposed approach and the ideal case for various BS antenna beam width and MSRs values.
It can be seen that when the BS beam becomes sharper, the gap of the achieved network utility between the proposed algorithm and the ideal case shrinks. As an extreme case where  $\Delta\theta^{\rm BS}={\pi}/{36}$ and $D^{\rm BS}= 40 \textrm{ dB}$, the proposed algorithm achieves almost identical performance to the ideal case, demonstrating its advantage.

\subsubsection{The Effect of Feedback Overhead} 
\begin{figure}[ht]
\centering
\includegraphics[width=0.48\textwidth]{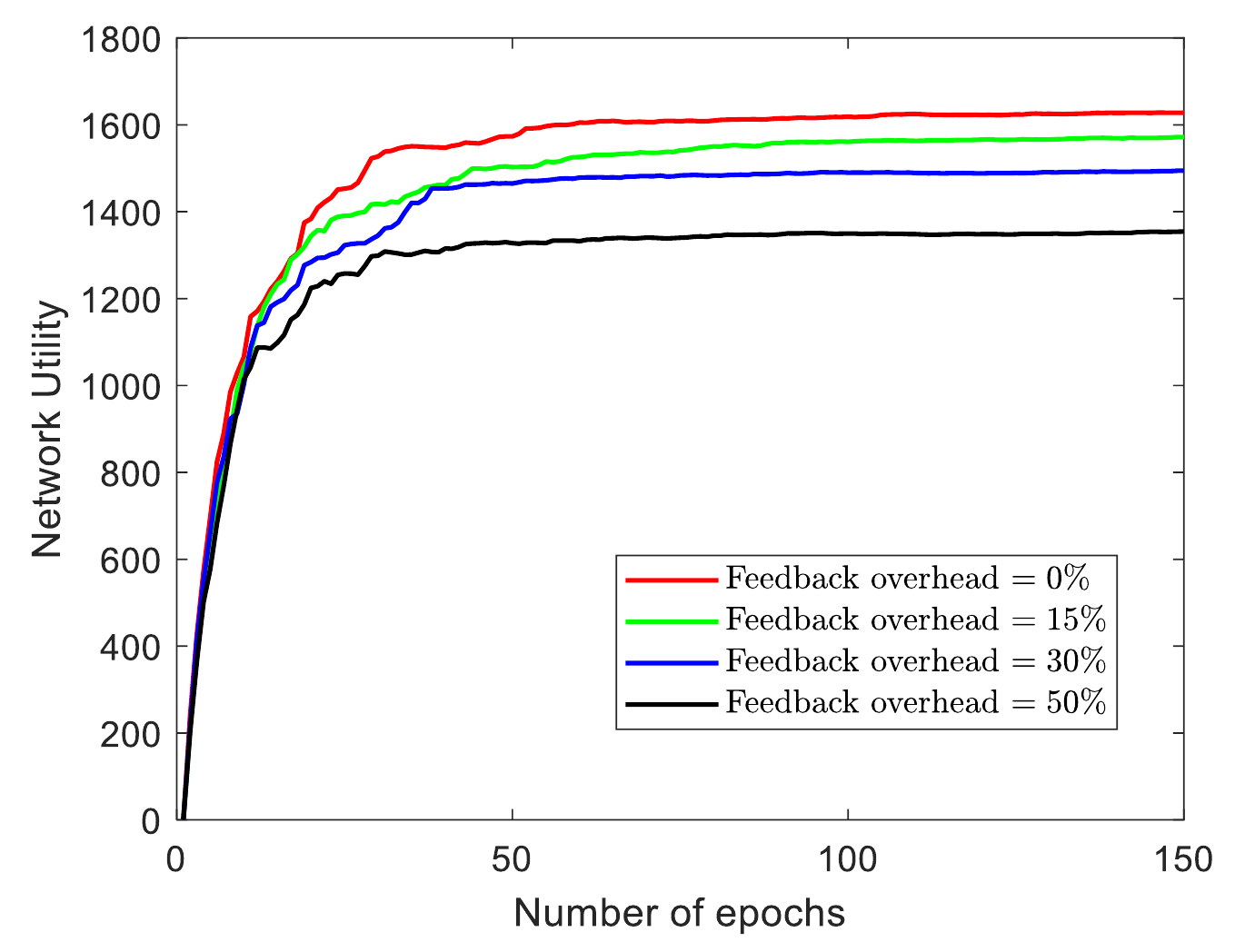}
\vspace{-0.2cm}
\caption{\small The effect of feedback overhead on  the performance of the proposed approach.}
\label{Fig: feedback overhead}
\end{figure}
In a time division system, a certain number of time slots need to be designated to the feedback process during which the selected UEs report the measured interference to the associated UEs. We verify the effect of the feedback overhead in this section. For the simulation, we further divide each time slot into multiple sub-slots, say 20 sub-slots per slot. We assign the first $S$ sub-slots of each slot for feedback. The reason for this further division is that, the size of the feedback message is usually very small because the measured interference value has to be quantized and possibly compressed and therefore devoting full slots to feedback may cause under-utilization of the resources. The BSs stay silent during the first $S$ sub-slots and then use the updated power starting from the $(S+1)^{\rm th}$ sub-slot. The {\it feedback overhead} is defined as $ 5\times S\%$, which is the portion of a slot that is used for feedback. We set the BS/UE beam width and MSR to be $\Delta\theta^{\rm BS} =\Delta\theta^{\rm UE}=\frac{\pi}{18}$, $D^{\rm BS}=20\textrm{ dB}$, $D^{\rm UE}=10\textrm{ dB}$. Fig.~\ref{Fig: feedback overhead} shows the result for various feedback overhead values. It can be seen that, first, the utility curve will converge slower as the BSs update their powers less frequently; second, the asymptotic network utility will drop since the average throughput decreases due to the non-transmission of the BSs during the feedback slots.

\section{Conclusion}
\label{sec: Conclusion}
In this work, we studied the distributed beam scheduling problem in 5G mm-Wave cellular networks where there is no cooperation or centralized coordination among base stations belonging to different operators that share the same spectrum. We proposed a new design framework based on the Lyapunov stochastic optimization techniques to maximize the network utility as a function of the time averaged throughput subject to the average and peak power constraints of the base stations. The original network utility optimization problem  was then transformed into two sub-problems which solve the auxiliary variables (convex) and the power allocation in each epoch (non-convex). With theoretical performance guarantees, we proposed a distributed beam scheduling algorithm to cope with the non-convexity of the second sub-problem by formulating the scheduling problem as a non-cooperative game where the optimal pricing factors determined by the virtual queues. 
An iterative interference-measuring based power update algorithm was proposed to solve the Nash Equilibrium and was shown to have fast converge speed. We numerically evaluated the effectiveness of the proposed scheduling algorithm compared to several baseline MAC scheduling algorithms including $p$-persistent and CSMA/CA protocols. The proposed optimization framework can accommodate a large range of other MAC protocols for network utility maximization, which opens up opportunities for future research. Furthermore, one may consider the scenario in which subsets of the base stations can cooperate and cooperative game-based approach may be investigated.

\section*{Acknowledgment}
This work 
was supported through the Idaho National Laboratory (INL) Laboratory Directed Research and Development (LDRD) Program under DOE Idaho Operations Office Contract DE-AC07-05ID14517 and the National Science Foundation grant SpecEES-1824558.

\appendices
\section{Proof of Lemma \ref{Lemma: BR}}
\label{Appendix: Proof of Lemma 1}
Observing that the payoff function $\phi_i$ is concave w.r.t. $p_{j(i),i}$, setting the first-order derivative to be zero, i.e., $\frac{\partial \phi_i(\pv_i,\pv_{-i})}{\partial p_{j(i),i}}=0$, we obtain
\begin{subequations}
\begin{align}
\frac{\partial \phi_i(\pv_i,\pv_{-i})}{\partial p_{j(i),i}}&=\frac{\alpha_iW}{1+ \textrm{SINR}_{j(i),i}}\frac{\partial \textrm{SINR}_{j(i),i}}{\partial p_{j(i),i}}-\lambda_i=0,\\
&\Rightarrow p_{j(i),i}=\frac{\alpha_iW}{\lambda_i}-\frac{1}{g_{j(i),i}} .
\end{align}
\end{subequations}
Considering the fact that the power can not be negative and the peak power constraint $p_i^{\rm max}$, we obtain
\be
p_{j(i),i}^{\rm BR}=\left[\frac{\alpha_iW}{\lambda_i}-\frac{1}{g_{j(i),i}} \right]_0^{p_i^{\rm max}}. 
\ee
which completes the proof of Lemma \ref{Lemma: BR}.

\section{Proof of Lemma 2}
\label{appendix: proof of lemma 2}
According to \cite{osborne1994course, aubin2007mathematical}, to prove the existence of NE for $\mathcal{G}$, we need to prove that: \emph{1)} The action space $\mathcal{P}_i$ for each BS $i$ is a nonempty, compact and convex subset of the Euclidean space $\mathbb{R}^{K_i}$; and \emph{2)} The payoff function $\phi_i$ for each BS $i$ is continuous on the product space $\prod_{i=1}^M\mathcal{P}_i\triangleq \mathcal{P}_1\times \mathcal{P}_2\times \cdots \times  \mathcal{P}_M$ and is quasi-concave on $\mathcal{P}_i$.

The NE of $\mathcal{G}$ indeed exists since: \emph{1)} The action space $\mathcal{P}_i$ consists of the set of all admissible power allocations for BS $i$ as shown in (\ref{eq: action space}). Since  the transmit power $p_{j,i}$ for each UE $ j\in\mathcal{K}_i$ is in a  closed interval $[0,p_i^{\rm max}]$, the action space $\mathcal{P}_i$ is compact (closed and bounded) and also convex;  \emph{2)}
 From (\ref{Eq: general GT payoff function}), it can be easily seen that the payoff function $\phi_i$ of each BS $i$ is continuous on the product space $\prod_{i=1}^M\mathcal{P}_i$ and is strictly concave w.r.t. $\pv_i$ on $\mathcal{P}_i$, which is a more strict condition than quasi-concavity.  This completes the proof of Lemma \ref{Lemma: existence of NE}.

\section{Proof of Theorem \ref{Theorem: Sufficient Conditions on the Uniqueness of NE}}
\label{Appendix: proof of Thm sufficient conditions}
We adopt a similar approach to that of \cite{pang2010design}. Starting from the assumption that $\mathbf{Q}$ is a P-matrix, we then prove that the mapping $\mathbf{F}$ is a uniformly P-function. More specifically, consider two power allocation profiles $\pv=(\pv_i)_{i=1}^M$ and $\pv'=(\pv_i')_{i=1}^M$ in $\mathcal{P}$. Note that although these two power allocation profiles might be different, the selected UE stays the same for each BS. From (\ref{Eq: VI mapping}), recalling that $\hbar_{j,i}\triangleq \sqrt{G^{\rm UE}_{j,i}G^{\rm BS}_{j,i}|h_{j,i}|^2d_{j,i}^{-\eta}}$,  we can write
\begin{subequations}
\begin{align}
&\mathbf{F}_i(\pv)=   \left[ \mathbf{0}^{\nu(i)-1}, \lambda_i-
\frac{\alpha_iW}{\sum_{i'\in[M]}|\widehat{\hbar}_{j(i),i'}|^2p_{j(i'),i'}+ \widehat{\sigma}_i^2}
,\mathbf{0}^{{K}/{M}-\nu(i)}   \right]^{\rm T},\\
&\mathbf{F}_i(\pv')=   \left[ \mathbf{0}^{\nu(i)-1}, \lambda_i-
\frac{\alpha_iW}{\sum_{i'\in[M]}|\widehat{\hbar}_{j(i),i'}|^2p_{j(i'),i'}'+ \widehat{\sigma}_i^2}
,\mathbf{0}^{{K}/{M}-\nu(i)}   \right]^{\rm T},
\end{align}
\end{subequations}
in which $\widehat{\hbar}_{j(i),i'}=\frac{\hbar_{j(i),i'}}{\hbar_{j(i),i}}$, $\widehat{\sigma}_i^2\triangleq\frac{\sigma^2}{|\hbar_{j(i),i}|^2},\forall i\in[M]$. Denote $\forall i\in[M]$: 
\be 
\varphi_i\triangleq\sqrt{ \sum_{i'\in[M]}|\widehat{\hbar}_{j(i),i'}|^2p_{j(i'),i'}'+ \widehat{\sigma}_i^2  }\sqrt{ \sum_{i'\in[M]}|\widehat{\hbar}_{j(i),i'}|^2p_{j(i'),i'}+ \widehat{\sigma}_i^2  },
\ee
and $\varphi_i^{\rm max}\triangleq\sum_{i'\in[M]}|\widehat{\hbar}_{j(i),i'}|^2p_{i'}^{\rm max}+ \widehat{\sigma}_i^2 $. It is easy to see that $\widehat{\sigma}_i^2 \le \varphi_i\le \varphi_i^{\rm max}$. Also denote $e_i\triangleq\frac{p_{j(i),i} -p_{j(i),i}'}{\varphi_i},\forall i\in[M]$. We then have
\begin{subequations}
\begin{align}
\frac{\left(\pv_i -\pv_i'\right)^{\rm T}\left(\mathbf{F}_i(\pv)-\mathbf{F}_i(\pv')   \right)}{\alpha_iW}&=\nonumber\\
\left(p_{j(i),i} -p_{j(i),i}'\right)&\left(   \frac{1}{ \sum_{i'\in[M]}|\widehat{\hbar}_{j(i),i'}|^2p_{j(i'),i'}'+ \widehat{\sigma}_i^2   }-\frac{1}{ \sum_{i'\in[M]}|\widehat{\hbar}_{j(i),i'}|^2p_{j(i'),i'}+ \widehat{\sigma}_i^2   }   \right)\\
&= \left(p_{j(i),i} -p_{j(i),i}'\right)\frac{ \sum_{i'\in[M]}|\widehat{\hbar}_{j(i),i'}|^2(p_{j(i'),i'}-p_{j(i'),i'}') }{\varphi_i^2} 
\end{align}
\begin{align}
&= e_i\left(e_i +  \frac{\sum_{i'\in [M]:i'\ne i} |\widehat{\hbar}_{j(i),i'}|^2\varphi_{i'}}{\varphi_i}    e_{i'}         \right)  \label{Eq: step 1}\\
&\ge e_i^2 - \frac{\sum_{i'\in [M]:i'\ne i} |\widehat{\hbar}_{j(i),i'}|^2\varphi_{i'}}{\varphi_i}|e_i||e_{i'}|\\
&\ge  |e_i|\left(\sum_{i'\in[M]}\frac{Q_{i,i'}}{\alpha_iW}   |e_{i'}|\right),\label{Eq: step 2}
\end{align}
\end{subequations} 
where $Q_{i,i'}$ is given by (\ref{Eq: Q matrix}); in (\ref{Eq: step 1}) we used the fact that $|\widehat{\hbar}_{j(i),i}|^2=1$ and (\ref{Eq: step 2}) is due to the fact that if $i'\ne i$, 
\be 
\frac{Q_{i,i'}}{\alpha_i W}=-\left|\frac{\hbar_{j(i),i'}}{\hbar_{j(i'),i'}}\right|^2\left(   1+ \frac{ \sum_{i''\in[M]}|\hbar_{j(i'),i''}|^2p_{i''}^{\rm max} }{\sigma^2} \right) \le -\frac{ |\widehat{\hbar}_{j(i),i'}|^2\varphi_{i'}}{\varphi_i},
\ee
which is because
\begin{subequations}
\begin{align}
\frac{Q_{i,i'}}{\alpha_i W}&=-\left|\frac{\hbar_{j(i),i'}}{\hbar_{j(i'),i'}}\right|^2 \frac{\sigma^2+ \sum_{i''\in[M]}|\hbar_{j(i'),i''}|^2p_{i''}^{\rm max}}{\sigma^2}    \\
&= -\frac{ \left|\hbar_{j(i),i'}\right|^2\left( \widehat{\sigma}_{i'}^2+ \sum_{i''\in[M]}|\widehat{\hbar}_{j(i'),i''}|^2p_{i''}^{\rm max}\right)}{\sigma^2}\\
&=-\frac{ \left|\hbar_{j(i),i'}\right|^2\varphi_{i'}^{\rm max}}{\sigma^2}
\le -\frac{ \left|\hbar_{j(i),i'}\right|^2\varphi_{i'}}{\sigma^2}\\
&= -\frac{ |\widehat{\hbar}_{j(i),i'}|^2\varphi_{i'}}{\widehat{\sigma}_{i}^2} \le -\frac{ |\widehat{\hbar}_{j(i),i'}|^2\varphi_{i'}}{\varphi_i},
\end{align}
\end{subequations}
throughout which we used the inequality $\widehat{\sigma}_i^2 \le \varphi_i\le \varphi_i^{\rm max}$. As a result, we have
\be
\label{Eq: ref 1}
\left(\pv_i -\pv_i'\right)^{\rm T}\left(\mathbf{F}_i(\pv)-\mathbf{F}_i(\pv')   \right) \ge |e_i|\left(\sum_{i'\in[M]}Q_{i,i'}|e_{i'}|\right).
\ee
Since we have assumed that $\mathbf{Q}$ is a P-matrix, we can  obtain a constant $C(\mathbf{Q})>0$ according to \cite{cottle2009linear} as follows:
\be
\label{Eq: P-matrix constant}
C(\mathbf{Q})\triangleq\min_{\|\xv\|_2^2=1}\left\{\max_{1\le i\le M}\left\{x_i\left(\mathbf{Q}\xv \right)_i\right\}\right\}>0,
\ee  
in which $\xv\triangleq(x_i)_{i=1}^M\in\mathbb{R}^{M}$ and $\left(\mathbf{Q}\xv \right)_i$ denotes the $i$-th entry of the vector $\mathbf{Q}\xv $. Scaling the vector $\xv$, we obtain an equivalent characterization as 
\be 
\label{Eq: ref 2} 
\max_{1\le i\le M}x_i\left(\mathbf{Q}\xv \right)_i \ge C(\mathbf{Q})\|\xv\|_2^2,\quad \forall \xv\in\mathbb{R}^{M}.
\ee
Combing (\ref{Eq: ref 1}) and (\ref{Eq: ref 2}), denoting $\yv\triangleq(|e_i|)_{i=1}^{M}$, we have
\begin{subequations}
\begin{align}
\max_{1\le i\le M} \left(\pv_i -\pv_i'\right)^{\rm T}\left(\mathbf{F}_i(\pv)-\mathbf{F}_i(\pv')   \right) &\ge 
\max_{1\le i\le M}y_i\left(\mathbf{Q}\yv \right)_i \ge   C(\mathbf{Q})\|\yv\|_2^2 = C(\mathbf{Q})\left(\sum_{i\in[M]}e_i^2 \right)\\
&\ge \frac{ C(\mathbf{Q})}{\max_{i\in[M]}\{\varphi_i^{\rm max}\}}\left( \sum_{i\in[M]}\left(p_{j(i),i} -p_{j(i),i}' \right)^2    \right)\\
& = \frac{ C(\mathbf{Q})}{\max_{i\in[M]}\{\varphi_i^{\rm max}\}}\|\pv-\pv'\|_2^2\\ 
&= C^{\rm up}\|\pv-\pv'\|_2^2,
\end{align}
\end{subequations}
in which the constant $C^{\rm up}\triangleq\frac{ C(\mathbf{Q})}{\max_{i\in[M]}\{\varphi_i^{\rm max}\}}$ does not depend on the power allocations. As a result, we obtain 
\be 
\max_{i\in[M]} \left(\pv_i -\pv_i'\right)^{\rm T}\left(\mathbf{F}_i(\pv)-\mathbf{F}_i(\pv')   \right) \ge C^{\rm up}\|\pv-\pv'\|_2^2\,,
\ee
proving that $\mathbf{F}$ is a uniformly P-function. According to Proposition \ref{Proposition: unique solution of VI(P,F)}, we conclude that the game $\mathcal{G}$ admits a unique NE.

\section{Proof of Lemma \ref{lemma: GT optimality gap}}
\label{appendix: proof of lemma 3}
Let $\bar X_{j,i}^{\rm ideal}$ denote the achieved average throughout of UE $j$ (of BS $i$) under the `ideal case' described in Section~\ref{Section: ideal case}. Also, let $\bar{X}_{j,i}^{\rm sub-opt}(\forall i\in[M],\forall j\in\mathcal{K}_i)$ be the optimal average throughput achieved by solving the two sub-problems (\ref{Eq: sub-opt 1}), (\ref{Eq: sub-opt 2}) at each epoch.  Since in the ideal case we assumed that there is no interference among BSs, the network utility produced by the ideal case is no worse than the utility achieved by optimally solving the two sub-problems at each epoch, i.e., $\sum_{i\in[M]}\sum_{j\in\mathcal{K}_i}U( \bar X_{j,i}^{\rm ideal}) \ge \sum_{i\in[M]}\sum_{j\in\mathcal{K}_i}U( \bar X_{j,i}^{\rm sub-opt})$. Therefore, the proposed scheduling algorithm is a $C$-approximation to the sub-problems. By \cite{neely2010stochastic} (Theorem 4.8), we obtain the optimality gap (\ref{eq: GT opt gap}) of the proposed scheduling algorithm.

\section{Optimality Gap Analysis}
\label{appendix: gap C analysis}
We consider the optimality gap between the non-cooperative game based solution to the second sub-problem and the ideal case. In both schemes, each BS $i$ randomly selects one of its associated UEs to transmit throughout the whole epoch, we let $j(i)\in\mathcal{K}_i$ denote the UE selected by BS $i,\forall i\in[M]$. Since we have assumed that the channel stays unchanged within each epoch, solving the second sub-problem can be decomposed into solving the following optimization problem at each block (the block and epoch indices are omitted here):
\begin{subequations}
\begin{align}
\max \quad &  \sum_{i\in[M]} \alpha_iW\log\left(1+\textrm{SINR}_{j(i),i}  \right) -\lambda_ip_{j(i),i}  \\
\textrm{s.t.} \quad & 0\le   p_{j(i),i}\le p_i^{\rm max},\quad \forall i\in[M].
\end{align}
\end{subequations}
in which $\alpha_i = H_{j(i),i}T^{\rm b},\lambda_i=Z_iT^{\rm b}$ ($H_{j(i),i}$ and $Z_i$ are the virtual queue status at the current epoch). We now compare the game based solution and the ideal case solution.

\subsubsection{Game based Solution} In the proposed scheduling approach, all BSs transmit to their selected UEs simultaneously so there is interference among them. Each BS $i$ selfishly maximizes its own payoff $\alpha_i W\log\left(1+\textrm{SINR}_{j(i),i}  \right) -\lambda_i p_{j(i),i}$ and we use the  Nash Equilibrium pf the game as an approximate solution to above optimization problem. Let $\pv^{\star}=[p_{j(1),1}^{\star},p_{j(2),2}^{\star},\cdots,p_{j(M),M}^{\star}]$ denote the equilibrium power allocation, then $\pv^{\star}$ satisfies 
\be
\label{eq: game based solution-appen}
p_{j(i),i}^{\star}  = \left[ \frac{\alpha_i W}{\lambda_i} -\frac{1}{g_{j(i),i}}    \right]_0^{p_i^{\rm max}},\quad \forall i\in[M].
\ee 
in which $g_{j(i),i} = \frac{G_{j(i),i}^{\rm UE}G_{j(i),i}^{\rm BS}|h_{j(i),i}|^2d_{j(i),i}^{-\eta}}{ \sum_{\ell\in[M]\backslash\{i\}} G_{j(i),\ell}^{\rm UE}G_{j(i),\ell}^{\rm BS}|h_{j(i),\ell}|^2d_{j(i),\ell}^{-\eta}p_{j(\ell),\ell}^{\star} +\sigma^2 }$ is the equivalent channel gain from BS $i$ to UE $j(i)$.

\subsubsection{Ideal Case Solution} In the ideal case, it is assumed there is no interference among BSs even though the BSs are transmitting simultaneously. Therefore, each BS $i$ only needs to maximize $\alpha_iW\log\left(1+\textrm{SNR}_{j(i),i}  \right) -\lambda_ip_{j(i),i}$ subject to the peak power constraint $p_{j(i),i}\le p_i^{\rm max}$. Let $\pv^{\rm ideal}=[p_{j(1),1}^{\rm ideal},\cdots,p_{j(M),M}^{\rm ideal}]$ denote the optimal solution then 
\be
\label{eq: ideal case solution-appen}
p_{j(i),i}^{\rm ideal} = \left[  \frac{\alpha_iW}{\lambda_i} -\frac{1}{g_{j(i),i}} \right]_0^{p_i^{\rm max}},\quad \forall i\in[M]. 
\ee
in which $g_{j(i),i}=\frac{G_{j(i),i}^{\rm UE}G_{j(i),i}^{\rm BS}|h_{j(i),i}|^2d_{j(i),i}^{-\eta}}{ \sigma^2 }$ is the equivalent channel gain from BS $i$ to UE $j(i)$.

Let $X_{j(i),i}(\pv^{\star})\triangleq TW\log(1+g_{j(i),i}p_{j(i),i}^{\star}) $  denote the throughput  achieved by the game based approach during an epoch and $X_{j(i),i}(\pv^{\rm ideal})\triangleq TW\log(1+g_{j(i),i}p_{j(i),i}^{\rm ideal}) $  denote the throughput  achieved by the ideal case during an epoch. Since there might be multiple  NEs, the utility gap $C$ between the game based solution and the ideal case can be determined by 
\be
C\ge \arg\max\limits_{\pv^{\star}\in\mathcal{E}} \left\{     
\sum_{i\in[M]}\sum_{j\in\mathcal{K}_i}U(X_{j,i}(\pv^{\rm ideal}))   -  \sum_{i\in[M]}\sum_{j\in\mathcal{K}_i}U(X_{j,i}(\pv^{\star}))    \right\},
\ee 
in which $\mathcal{E}$ is the set of NEs of the game. Note that $X_{j,i}(\pv^{\rm ideal})=X_{j,i}(\pv^{\star})=0, \forall j\in \mathcal{K}_i\backslash\{j(i)\},\forall i\in[M]$, i.e., non-selected UEs have zero throughput.

\bibliographystyle{IEEEtran}
\bibliography{references_d2d}

\end{document}

%% file: macros.tex
\long\def\comment#1{}

\newfont{\bbb}{msbm10 scaled 700}

\newfont{\bb}{msbm10 scaled 1100}

\newcommand{\EE}{\mbox{\bb E}}

\newcommand{\pv}{{\bf p}}

\newcommand{\xv}{{\bf x}}
\newcommand{\yv}{{\bf y}}

\newcommand{\Gc}{{\cal G}}

\renewcommand{\arg}{{\hbox{arg}}}

\newcommand{\eqdef}{\stackrel{\Delta}{=}}

\newcommand{\be}{\begin{equation}}
\newcommand{\ee}{\end{equation}}
\newcommand{\bea}{\begin{eqnarray}}
\newcommand{\eea}{\end{eqnarray}}

